\documentclass[runningheads]{llncs} 

\usepackage{algorithm}
\usepackage[noend]{algpseudocode}
\usepackage[T1]{fontenc}
\usepackage{graphicx}
\usepackage{caption}
\algblock{Input}{EndInput}
\algnotext{EndInput}
\algblock{Output}{EndOutput}
\algnotext{EndOutput}
\newcommand{\Desc}[2]{\State \makebox[4em][l]{#1}#2}

\title{Reasoning about Rare-Event Reachability in Stochastic Vector Addition Systems via Affine Vector Spaces}
\titlerunning{Reasoning about Reachability in S-VAS via Vector Spaces}
\author{Joshua Jeppson\inst{1}\orcidID{0000-0002-4177-7604} 
\and
Landon Taylor\inst{1}\orcidID{0000-0002-4071-3625}
\and
Bingqing Hu\inst{1}\orcidID{0009-0006-5933-6141}
\and
Zhen Zhang\inst{1}\orcidID{0000-0002-8269-9489}}%
\authorrunning{J. Jeppson et al.}
\institute{Utah State University, Utah, USA}

\usepackage{amsmath,amssymb,amsfonts,mathtools}
\usepackage{hyperref}
\usepackage[pdftex,dvipsnames]{xcolor}
\usepackage{enumerate}
\usepackage{enumitem}
\usepackage{graphicx}
\usepackage{xargs}             
\usepackage{xspace}
\usepackage[normalem]{ulem}
\usepackage{booktabs} 
\usepackage{underscore} 
\usepackage{tabularx}
\usepackage{adjustbox}
\usepackage{wrapfig}
\usepackage{multirow, multicol}
\usepackage{booktabs}
\usepackage{multirow}
\usepackage{amsmath}
\usepackage{amssymb}
\newcolumntype{C}[1]{>{\centering\arraybackslash}m{#1}}
  
\newcommand{\stSetSize}{N}

\usepackage{colortbl} 

\errorcontextlines\maxdimen

\makeatletter
\newcommand*{\algrule}[1][\algorithmicindent]{\makebox[#1][l]{\hspace*{.5em}\vrule height .75\baselineskip depth .25\baselineskip}}%

\newcount\ALG@printindent@tempcnta
\def\ALG@printindent{%
    \ifnum \theALG@nested>0
        \ifx\ALG@text\ALG@x@notext
            \addvspace{-3pt}
        \else
            \unskip
            \ALG@printindent@tempcnta=1
            \loop
                \algrule[\csname ALG@ind@\the\ALG@printindent@tempcnta\endcsname]%
                \advance \ALG@printindent@tempcnta 1
            \ifnum \ALG@printindent@tempcnta<\numexpr\theALG@nested+1\relax
            \repeat
        \fi
    \fi
    }%
\usepackage{etoolbox}
\patchcmd{\ALG@doentity}{\noindent\hskip\ALG@tlm}{\ALG@printindent}{}{\errmessage{failed to patch}}
\makeatother

\newcommand{\nonNegInt}{\ensuremath{\mathbb{Z}_{\geqslant 0}}}

\newcommand{\real}{\ensuremath{{\mathbb{R}}}}
\newcommand{\posReal}{\ensuremath{{\mathbb{R}^{+}}}}
\newcommand{\nonNegReal}{\ensuremath{\mathbb{R}_{\geqslant 0}}}
\newcommand{\rational}{\ensuremath{\mathbb{Q}}}

\newcommand{\st}{\ensuremath{\vec{s}}\xspace}
\newcommand{\state}[1]{\ensuremath{\st_{#1}}\xspace}  
\newcommand{\nextSt}{\ensuremath{\st_{nxt}}\xspace} 
\newcommand{\absSt}{\ensuremath{\st_{abs}}\xspace}
\newcommand{\stSet}{\ensuremath{\mathcal{S}}\xspace} 
\newcommand{\initSt}{\ensuremath{\state{0}}\xspace} 

\newcommand{\trace}{\ensuremath{\sigma}\xspace} 

\newcommand{\ctmc}{\ensuremath{\mathtt{C}}} 
\newcommand{\ctmcFull}{\ensuremath{\ctmc = \langle \stSet, \initSt,
		\tranRateShort, \labelFunc \rangle}\xspace} 
\newcommand{\labelFunc}{\ensuremath{\mathbf{L}}} 

\newcommand{\tranRateShort}{\ensuremath{\mathbf{R}}\xspace} 
\newcommand{\tranRate}[2]{\ensuremath{\tranRateShort(#1, #2)}\xspace} 
\newcommand{\exitRate}[1]{\ensuremath{\mathnormal{E} (#1)}} 
\newcommand{\tranProb}[2]{\ensuremath{p(#1, #2)}} 

\newcommand{\true}{{\fontfamily{cmtt}\selectfont{true}}}

\newcommand{\propen}[1]{\ensuremath{\mathnormal{\theta_{#1}}}} 
\newcommand{\stChangeVec}[1]{\ensuremath{\vec{r}_{#1}}} 
\newcommand{\react}[1]{\ensuremath{\mathcal{R}_{#1}}} 
\newcommand{\reactDef}[1]{\ensuremath{\react{#1} = \langle
		\reactantVec{#1}, \productVec{#1}, \reactionRateConst{#1}, 
		\propen{#1} \rangle}} 
\newcommand{\reactSet}{\ensuremath{\mathfrak{R}}\xspace} 
\newcommand{\reactSetSize}{\ensuremath{\mathnormal{n}}\xspace} 
\newcommand{\reactSetFull}{\ensuremath{\reactSet = \{\react{0}, \dots, 
		\react{\reactSetSize-1}\}}\xspace} 
\newcommand{\crn}{\ensuremath{\mathcal{M}}\xspace}
\newcommand{\crnFull}{\ensuremath{\crn = \langle \speciesSet,
		\reactSet, \initSt \rangle}\xspace} 
\newcommand{\species}[1]{\ensuremath{\mathcal{X}_{#1}}} 
\newcommand{\speciesSet}{\ensuremath{\mathfrak{X}}}
\newcommand{\speciesSetSize}{\ensuremath{\mathnormal{m}}\xspace} 
\newcommand{\speciesSetFull}{\ensuremath{\speciesSet = [\species{1}, \dots,
		\species{\speciesSetSize}]^T}} 
\newcommand{\reactantSet}[1]{\ensuremath{\mathsf{Reactant}_{#1}}}

\newcommand{\reactantVec}[1]{\ensuremath{\mathbf{rv}_{#1}}} 
\newcommand{\productVec}[1]{\ensuremath{\mathbf{pv}_{#1}}} 
\newcommand{\reactionRateConst}[1]{\ensuremath{k_{#1}}} 

\newcommand{\probOp}{\ensuremath{\mathsf{P}}} 
\newcommand{\pMin}{\ensuremath{\mathsf{P}_{min}}\xspace} 
\newcommand{\pMax}{\ensuremath{\mathsf{P}_{max}}\xspace} 

\newcommand{\postUntil}{\ensuremath{\mathit{\Psi}}\xspace}

\newcommand{\upperTimeBound}{\ensuremath{T}} 
\newcommand{\eventuallyFull}{\ensuremath{\Diamond^{[0, \upperTimeBound]} \,
    \postUntil}}
\newcommand{\eventuallyFullUnbounded}{\ensuremath{\Diamond\,
		\postUntil}}
\newcommand{\cslEventually}{\ensuremath{\probOp_{=?}
    (\eventuallyFull)}}
\newcommand{\cslEventuallyUnbounded}{\ensuremath{\probOp_{=?} (\eventuallyFullUnbounded)}}

\newcommand{\depEdge}{\ensuremath{\leadsto}\xspace} 

\newcommand{\vas}{\ensuremath{\mathcal{V}}\xspace}

\newcommand{\ragtimer}{\textsc{Ragtimer}\xspace}
\newcommand{\prism}{\textsc{Prism}\xspace}
\newcommand{\storm}{Storm\xspace}
\newcommand{\stamina}{\textsc{Stamina}\xspace}
\newcommand{\modest}{\textsc{Modest}\xspace}
\newcommand{\modestToolset}{\modest\ \textsc{Toolset}\xspace}
\newcommand{\modes}{\texttt{modes}\xspace}
\newcommand{\isr}{\textsc{Isr}\xspace} 
\newcommand{\sdp}{\textsc{Sdp}\xspace} 

\newcommand{\bfred}[1]{\textsf{\bfseries\color{Maroon}#1}\xspace}
\newcommand{\bfblue}[1]{\textsf{\bfseries\color{BlueViolet}#1}\xspace}
\newcommand{\bfgreen}[1]{\textsf{\bfseries\color{OliveGreen}#1}\xspace}
\newcommand{\bfyellow}[1]{\textsf{\bfseries\color{Bittersweet}#1}\xspace}

\usepackage[]{todonotes}

\newcommandx{\jj}[2][1=]{\todo[inline,linecolor=LimeGreen,backgroundcolor=SpringGreen!40,bordercolor=LimeGreen,#1]{\bfgreen{(Josh) #2}}}
\newcommandx{\jjzz}[2][1=]
{\todo[inline,linecolor=LimeGreen,backgroundcolor=SpringGreen!40,bordercolor=LimeGreen,#1]{\bfgreen{(Josh) From Zhen: #2}}}

\newcommandx{\lt}[2][1=]{\todo[inline,
  linecolor=RawSienna,backgroundcolor=Salmon!40,bordercolor=RawSienna,#1]{\bfred{(Landon) #2}}}  

\newcommandx{\bh}[2][1=]{\todo[inline, linecolor=Cerulean, 
backgroundcolor=Cerulean!25, bordercolor=Cerulean]{\bfblue{(Beckey) #2}}}

\newcommandx{\zz}[2][1=]{\todo[inline,
  linecolor=MidnightBlue,backgroundcolor=CornflowerBlue!40,bordercolor=MidnightBlue,#1]{\bfblue{(Zhen)
  #2}}}

\newcommandx{\all}[2][1=]{\todo[inline,
linecolor=Bittersweet,backgroundcolor=Bittersweet!40,bordercolor=Bittersweet,#1]{\bfyellow{(All) #2}}}

\newcommandx{\rc}[2][1=]{\todo[inline,
linecolor=Bittersweet,backgroundcolor=Bittersweet!40,bordercolor=Bittersweet,#1]{\bfyellow{(Reviewer) #2}}}

\newcommand{\projMat}{\ensuremath{\mathcal{P}}\xspace}
\newcommand{\residual}{\ensuremath{\vec{\epsilon}}\xspace}
\newcommand{\residmag}{\ensuremath{\epsilon}\xspace}
\newcommand{\mld}{\ensuremath{\text{\textup{\texttt{mld}}}}\xspace}

\newcommand{\Sp}{\ensuremath{\mathsf{S}}\xspace}

\newcommand{\solutionSpace}{\ensuremath{V_{\postUntil}}\xspace}
\newcommand{\reachableSpace}{\ensuremath{V_\vas}\xspace}
\newcommand{\spaceMapping}{\mathtt{m}\xspace}

\newcommand{\foundSatStates}{\ensuremath{\mathtt{F}_{\postUntil}}\xspace}
\newcommand{\subspacesCount}{\ensuremath{\mu}\xspace} 

\begin{document}

\maketitle
\begin{abstract}
Rare events in Stochastic Vector Addition System (VAS) are of significant interest because, while extremely unlikely, they may represent undesirable behavior that can have adverse effects. Their low probabilities and potentially extremely large state spaces challenge existing probabilistic model checking and stochastic rare-event simulation techniques.
In particular, in Chemical Reaction Networks (CRNs), a chemical kinetic language often represented as VAS, rare event effects may be pathological.
We present two novel heuristics for priority-first partial state space expansion and trace generation tuned to the transient analysis of rare-event probability in VAS: Iterative Subspace Reduction (\isr) and Single Distance Priority (\sdp). Both methods construct a closed vector space containing all solution states.
\sdp then simply prioritizes shorter distances to this ``solution space'', while
\isr constructs a set of nested subspaces, 
where short and highly-probable satisfying traces are likely 
to pass through in sequence. 
The resulting partial state graph from each method contains likely traces to rare-event states, allowing efficient probabilistic model checking to compute a lower-bound probability of a rare event of interest.
These methods are deterministic, fast, and demonstrate marked performance on challenging CRN models.

\end{abstract}

\section{Introduction}

Quantitative analysis of \textit{Continuous-time Markov chains} (CTMCs) is critical to understanding real-world phenomena, including game state machines, economics, population models, and synthetic biology. \textit{Chemical Reaction Networks} (CRNs) are of particular interest, as they are a general-purpose language to model the continuous-time probabilistic chemical kinetics in biochemical reaction networks~\cite{Chellaboina2009}, genetic regulatory networks~\cite{Myers2009}, and molecular programming~\cite{Soloveichik2010}.
A CRN is represented as a Stochastic \emph{Petri Net}~\cite{leroux2012}, or equivalently a \emph{Vector Addition System} (VAS). Even a simple VAS can incur intractible state counts.

CRNs often operate in noisy environments where irrelevant inputs exponentially increase the number of system configurations. 
Although extremely infrequent, rare events in CRNs are of interest as they reveal undesirable behaviors that can lead to detrimental pathological consequences. Thus, it is crucial to effectively obtain precise quantitative reliability guarantees for rare events.

\textit{Probabilistic model checking} (PMC) can provide provable guarantees of the transient time-bounded probability of a rare event. However, its computation requires the generation of the complete explicit state space, limiting its scalability and practicality. Efficient PMC tools~\cite{Jeppson2023,Roberts2022,Neupane2019,Hahn2009} construct a partial state graph to obtain probability bounds, but they still face a stiff scalability barrier for rare events. In contrast, stochastic simulation techniques~\cite{Budde2019,Kuwahara2008,Donovan2013} often require manual efforts to guide simulation toward a rare event. Recent work~\cite{Taylor2023,Israelsen2023} enumerates a large set of traces to obtain a lower probability bound, but it relies on randomized trace generation that may miss high-probability traces and suffer degraded performance due to duplicated traces. We present scalable and efficient solutions to enable PMC for transient CTMC numerical analysis on models that overwhelm these existing state-of-the-art PMC tools.

\textit{In this paper, we present
\emph{Iterative Subspace Reduction} (\isr) and \emph{Single Distance Priority} (\sdp),
two novel heuristics for a priority-first state-search algorithm to efficiently build a partial state space for PMC.}
We leverage linear algebraic techniques to exclude low-probability states and bias toward states that represent a rare event of interest. Unlike existing rare-event heuristics, which require the user to have deep knowledge of the model, and rely on nondeterministic sampling, the presented techniques \emph{automatically} create a solution vector space that contains \emph{all} satisfying states.
\isr uses information from a dependency graph~\cite{Israelsen2023} 
to create a set of nested subspaces around reachable rare-event states, whereas \sdp simply compares states' distances to the solution space. 
Both \isr and \sdp perform exceptionally well on complex models that pose a significant challenge to traditional PMC tools, 
as well as cutting-edge state-space truncation PMC tools.
Our results show that \isr appears more scalable, though it is outperformed on some models by \sdp.
To the best of our knowledge, this is the first approach to apply affine vector space techniques to tackle the challenging \emph{quantitative} VAS reachability problem. 

The remainder of the paper is organized as follows. Section~\ref{sec:prelims} details background information for this work, and Section~\ref{sec:related} describes related work. Section~\ref{sec:algorithm} details the primary contribution of this paper, i.e., the theoretical foundations and algorithm for \isr and \sdp. Section~\ref{sec:implementation} describes our implementation, and Section~\ref{sec:results} describes our empirical findings.

\section{Preliminaries}\label{sec:prelims}

\paragraph{Notation.} Generally, this paper uses standard mathematical notation. $\nonNegInt^m$ indicates the set of all nonnegative integer vectors of length $m$,
\noindent while $\nonNegInt^{<p}$ indicates the set of integers in $[0, p)$. A vector $\vec{x}$ is denoted in bold and $\vec{x}[i]$ indicates the $i^{th}$ element of $\vec{x}$. For a set or sequence, $|\trace|$ indicates \emph{cardinality}, or the number of elements of \trace, while $|\vec{x}|_1$ or $|\vec{x}|$ indicates the L1 norm of vector $\vec{x}$.
\paragraph{Continuous-Time Markov Chain (CTMC) and property specification.}
A CRN induces a \emph{continuous-time Markov chain} (CTMC). Formally, a CTMC is a tuple \ctmcFull,  where \stSet is the set of reachable states called the \emph{state 
space} (or \emph{state graph}); \ensuremath{\initSt \in \stSet} is the initial state; 
\ensuremath{\tranRateShort : \stSet \times \stSet \rightarrow \nonNegReal} 
is the transition rate matrix; and \ensuremath{\labelFunc : \stSet \rightarrow 
2^{AP}} is a state labeling function with atomic proposition set \ensuremath{AP}. 
A CRN reaction \react{i} is \emph{enabled} to fire in a state \st if its \emph{reaction rate}, defined by propensity function \propen{i}(\st), evaluates to a positive value. 
The probability that \react{i} occurs 
is \ensuremath{\tranProb{\st}{\nextSt} = 
\frac{\tranRate{\st}{\nextSt}}{\exitRate{\st}}}, where the \emph{exit rate} 
$\exitRate{\st}$ 
is the sum of all enabled reactions from \st.
Thus, the probability of exiting a state \st\ 
in time interval \ensuremath{[0, t]} is 
\ensuremath{1 -e^{-\exitRate{\st} \cdot t}}, where $t \in 
\real_{\geqslant 0}$ represents real time.
This work provides a guaranteed lower bound
on the solution to the time-bounded transient reachability property \cslEventually, where \postUntil is a non-nested \emph{satisfying state formula}, specified in \emph{Continuous Stochastic Logic} (CSL)~\cite{Aziz2000,Kwiatkowska2007}. This formula asks ``what is the probability that a model will reach a satisfying state \postUntil within $T$ time units?'' and may be checked via PMC. In CRNs, \postUntil can encode \emph{rare events}, whose probability is extremely low. Rare events can be undesirable states whose probabilities should not exceed a threshold. 
In a large population of cells, events that are rare in a single cell become more likely due to the sample size.

If the entire state space is not known, an artificial \emph{absorbing state}, \absSt, can be introduced to represent all unexplored states. If $\absSt \nvDash \postUntil$, the PMC result of a \emph{partial state graph} represents a lower bound on the actual value of \cslEventually.


\paragraph{Vector Addition System (VAS) and Reachability.}
A VAS~\cite{Karp1969,Lipton1976} (equivalently a Petri Net~\cite{Leroux2011})
is a finite set of vectors $\vas \subseteq \mathbb{Z}^m$. 
The size of \vas\ is denoted \reactSetSize, and the $i^{th}$ vector is $\stChangeVec{i}$. The initial state of \vas\ is denoted \initSt, and thus any reachable state $\st$ can be found by the equation $\st = R\vec{a} + \initSt$, where $\vec{a} \in \nonNegInt^\reactSetSize$ and $R$ is an $m \times \reactSetSize$ matrix whose column $i$ is \stChangeVec{i}. 
A trace \trace to \st is a unique sequence of \stChangeVec{i}'s and is \emph{valid} iff (1) for $\state{j} = \initSt + \sum_{j = 0}^k\trace[j]$ that $\state{j} \in \nonNegInt^m$ for \emph{all} $k \leqslant |\trace|$, that is, all intermediate \state{j} states remain in the first orthant; and (2) $\st = \state{|\trace|} = \initSt + \sum_{j = 0}^{|\trace|}\trace[j]$. A \trace predicted by the solution vector $\vec{a}$ will have $|\trace| = |\vec{a}|_1$, where the number of times $\stChangeVec{i}$ occurs in \trace is $\vec{a}[i]$.
State \st is \emph{reachable} iff there is 
a valid trace from \initSt to \st. A \emph{continuous time stochastic} Petri Net or VAS associates transition vector \stChangeVec{i} to \ensuremath{\propen{i} : \nonNegInt^{\speciesSetSize} \, \mapsto 
\posReal}, where \ensuremath{\propen{i}(\st)} represents the transition rate on VAS edge \ensuremath{\st \to \st + \stChangeVec{i}}. 
It is thus a CTMC, with \stSet being the set of all valid, reachable states, and $\tranRateShort$ is populated via the functions \ensuremath{\propen{i}, i \in \nonNegInt^\reactSetSize}. $AP$ is the set of all predicates on the $\speciesSetSize$ state variables, and $\labelFunc(\st)$ is the subset of $AP$ which evaluate to \true\ on \st.
In this paper, \emph{``VAS'' refers to stochastic VAS.}

%
Conventional reachability in VAS (e.g., \cite{Lipton1976,Rackoff1978,Czerwinski2020b}) asks the following question: for some state $\st_s$, 
does there exist a \emph{valid} trace of nonzero probability
$\trace \subseteq \vas$ such that $\initSt \xrightarrow{\trace} \st_s$ (i.e., $\trace$ goes from \initSt to $\st_s$)?
\noindent In this paper, we reason about the state formula \postUntil, where state $\st_s$ is \emph{satisfying} iff $\st_s \models \postUntil$.
Furthermore, our techniques target \emph{quantitative reachability} by identifying \emph{numerous} traces that lead to states that are both  \emph{reachable} and \emph{satisfying}, and by computing their aggregate probability within a given time bound.

\paragraph{Chemical Reaction Networks (CRNs).} CRNs are probabilistic and evolve in real-time. Formally, a CRN is a tuple \crnFull with \speciesSetSize chemical species arranged in a vector
\speciesSetFull, \reactSetSize reactions \reactSetFull, initial state vector
\ensuremath{\initSt : \speciesSet^{\speciesSetSize} \rightarrow	\nonNegInt} representing all species' initial molecule count, where \ensuremath{\speciesSetSize, \reactSetSize \in \nonNegInt} and 
\ensuremath{\speciesSetSize, \reactSetSize < \infty}. A reaction tuple \reactDef{i} is composed of a reactant vector \ensuremath{\reactantVec{i} \in \nonNegInt^{\speciesSetSize}} denoting 
the stoichiometry of reactants, a product vector 
\ensuremath{\productVec{i} \in \nonNegInt^{\speciesSetSize}} denoting the 
stoichiometry of products, a reaction rate coefficient \ensuremath{\reactionRateConst{i} \in \posReal}, 
and a \emph{propensity function} \ensuremath{\propen{i} : \nonNegInt^{\speciesSetSize} \, \mapsto 
\posReal}, which determines the probability of firing \react{i} in a state. 
\ensuremath{\propen{i}(\st)} is the product of \reactionRateConst{i} and the number of possible combinations of reactant molecules at state \st: \ensuremath{\propen{i} (\st) = \reactionRateConst{i}\prod_{\species{j} \in \reactantSet{i}} (\st[j])}. The \emph{update vector}, \ensuremath{\stChangeVec{i} = \productVec{i} - \reactantVec{i}}, represents the change of each species' molecule count from executing reaction \react{i}, e.g., for \react{i}: \ensuremath{\species{1} + \species{2} \to \species{3}}, its \emph{update vector} is $\begin{bmatrix}-1 & -1 & 1 \end{bmatrix}^T$, signifying consumption of \species{1} and \species{2} and production of \species{3}. All CRNs follow the \emph{Stochastic Chemical Kinetic} assumption~\cite{Myers2009}: Each element of \stChangeVec{i} is limited to \ensuremath{0, \pm1, \pm2} due to \react{i} occurring almost instantaneously, and each reaction can have at most three reactants.

\textit{Probabilistic Model Checking. }
PMC formally verifies models with probabilistic characteristics.
\prism~\cite{Kwiatkowska2011}, \textsc{Modest}~\cite{Hartmanns2014}, and 
\storm~\cite{Hensel2022} 
attempt to find accurate probabilities, but are limited by their requirement for \textit{explicit} state representation~\cite{Hensel2021} for transient numerical CTMC analysis, which causes state explosion when the state count is large or infinite.
\noindent Of the existing PMC tools, this work is comparable to \ragtimer~\cite{Israelsen2023,Taylor2023} and \stamina~\cite{Jeppson2023,Roberts2022,Neupane2019}. 
Both \stamina and our approach create partial state graphs \emph{on-the-fly}; however, like \ragtimer, we seek out traces leading to satisfying states, substantially reducing the state count. In contrast, \stamina searches for the most probable states, returning a probability bound $[\pMin, \probOp_{max}]$ such that the true probability lies within.
Unlike this work, \stamina is minimally aware of the property of interest during state expansion and thus performs worse on rare-event models.
\cite{Israelsen2023} demonstrates that many rare-event properties do not easily converge on an upper bound, 
\noindent causing \stamina to struggle terminating in a reasonable time. 

\paragraph{Motivating Example.}
The motivating example for this work is the \emph{modified yeast polarization}
model~\cite{Daigle2011}, a CRN with eight reactions and seven species:
\[
\resizebox{\textwidth}{!}{
$\displaystyle
\begin{array}{lll}
    \react{1} : \ \emptyset \xrightarrow{0.0038} \textrm{R}, &
    \react{2} : \ \textrm{R} \xrightarrow{4.00\times 10^{-4}} \emptyset,~~~ &
    \react{3} : \ \textrm{L} + \textrm{R} \xrightarrow{0.042} \textrm{RL} + \textrm{L}, \\
    \react{4} : \ \textrm{RL} \xrightarrow{0.010} \textrm{R},~~~ &
    \react{5} : \ \textrm{RL} + \textrm{G} \xrightarrow{0.011} \textrm{G}_\textrm{a} + \textrm{G}_{\textrm{bg}},~~~ &
    \react{6} : \ \textrm{G}_\textrm{a} \xrightarrow{0.100} \textrm{G}_\textrm{d}, \\
    \react{7} : \ \textrm{G}_\textrm{d} + \textrm{G}_{\textrm{bg}} \xrightarrow{1.05\times 10^{3}} \textrm{G},~~~ &
    \react{8} : \ \emptyset \xrightarrow{3.21} \textrm{RL}. & \\
\end{array}$
}
\]
Species vector \ensuremath{(R, L,  RL, G, G_{a}, G_{bg}, G_d)} is 
initially
\ensuremath{\initSt = [50,
  2, 0, 50, 0, 0, 0]^T}. 
This model is an adaptation of the pheremone-induced G-protein cycle in Saccharomyces cerevisia~\cite{Drawert2010}. A constant population of ligand ($\textrm{L}=2$) prevents this circuit from reaching equilibrium. The rare event of interest is a rapid (within 20 seconds) build-up from a molecule count of $0$ to $50$ for species $G_{bg}$, which causes the model to become \textit{polarized}, an undesirable behavior. Specifically, we check the CSL property \ensuremath{\probOp_{=?}(\Diamond^{[0, 20]} \, G_{bg}=50)}, querying the probability that within 20 seconds the molecule count of $G_{bg}$ increases to $50$.

For example, if \react{3} is executed from the given initial state in the 
motivating example to reach 
\ensuremath{\state{1} = [49, 2, 1, 50, 0, 0, 0]}, 
\reactantVec{5} and \productVec{5} for \react{5} are 
\ensuremath{[0, 0, 1, 1, 0, 0, 0]} and \ensuremath{[0, 0, 0, 0, 1, 1, 0]}, 
respectively; 
$\stChangeVec{5}$, the update vector for \react{5}, is $\stChangeVec{5} = \productVec{5} - \reactantVec{5}$; 
\reactionRateConst{5} is \ensuremath{0.011}; and 
\ensuremath{\propen{5} (\state{1}) 
=\reactionRateConst{5}(\state{1}[2])(\state{1}[3])= 0.011 \cdot 1 \cdot 50 = 
0.55 > 0}, so \react{5} is enabled in \state{1}.
Additional enabled reactions and their propensities at the initial state are 
\react{1} (\ensuremath{0.0038}), \react{2} (\ensuremath{0.0196}), \react{3} 
(\ensuremath{4.116}), \react{4} (\ensuremath{0.01}), and \react{8} 
(\ensuremath{3.21}).
The exit rate \ensuremath{\exitRate{\state{1}}} is 
\ensuremath{7.9094} and the probability that \react{5} executes 
is \ensuremath{0.55/7.9094 \approx 0.0695}. 
This model is highly concurrent, and it takes at least $100$ reaction executions to reach a satisfying state. While it appears simple at first, it is extremely challenging for state-of-the-art tools, detailed in Section~\ref{sec:results}.

\begin{wrapfigure}[9]{L}{0.33\linewidth}
\vspace{-20pt}
	\centering
	\includegraphics[width=\linewidth]{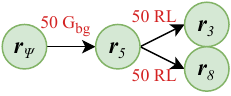}
	\caption{Dependency graph adapted from~\cite{Israelsen2023}.}
	\label{fig:depgraph}
\end{wrapfigure}

\paragraph{Dependency Graph.}
A transition \emph{dependency graph}~\cite{Israelsen2023} 
constructed from a VAS
and the satisfying state formula \postUntil 
is a directed acyclic graph containing as nodes the transitions required for satisfying state reachability 
and their dependencies.
For example, the dependency relation \ensuremath{\stChangeVec{i} \depEdge \stChangeVec{j}} 
is intuitively understood as \textit{\ensuremath{\stChangeVec{i}} depends on 
\ensuremath{\stChangeVec{j}}},
or more formally, $\exists\ p \leqslant m$ such that $\stChangeVec{j}[p] > 0 \implies \stChangeVec{i}[p] < 0$.
Dependency relations 
(edges in the dependency graph) are labeled with the name and quantity of the 
species that required the creation of the relation. Leaf nodes represent 
transitions that are enabled in the initial state. If a dependency graph cannot be constructed, satisfying states are proven non-reachable~\cite{Israelsen2023}. Graph 
construction begins with an abstract transition 
\ensuremath{\stChangeVec{\postUntil}} representing \postUntil.
Definition~\ref{def:deprelation} generalizes dependency relations beyond CRNs to 
VAS by defining enabledness and dependency in terms of vector quantities. 
\begin{definition} [Dependency Relation]
	\label{def:deprelation}
	A dependency graph is constructed from dependency relations 
	\ensuremath{\depEdge \subseteq \vas \times \vas} such that
	\ensuremath{\stChangeVec{i} \depEdge \stChangeVec{j}, i \neq j} if and only if:
        (1) \ensuremath{\stChangeVec{i}} is not enabled in the initial state (that is, $\initSt + \stChangeVec{i} \notin \nonNegInt^m$), and
		(2) \ensuremath{\lnot (\stChangeVec{j} \depEdge^{+} \stChangeVec{i}) }, and
		(3) $\exists\ p \leqslant m$ such that $\stChangeVec{j}[p] > 0 
		\implies \stChangeVec{i}[p] < 0$ (in CRNs, this means 
		\ensuremath{\react{i}} consumes a species produced by 
		\ensuremath{\react{j}}), and
		(4) \ensuremath{\stChangeVec{j}} is enabled in the initial 
		state, or \ensuremath{\exists \stChangeVec{k} \text{~such that~} 
		\stChangeVec{j} \depEdge \stChangeVec{k}}.
\end{definition}
Consider the motivating example. Figure~\ref{fig:depgraph} shows the dependency graph for this model. First, \react{\postUntil}, an 
artificial abstract transition, denotes that $G_{bg} = 50$ constitutes a satisfying state. The 
only reaction that produces $G_{bg}$ is \react{5}, so an edge to node 
\react{5}\ labeled $G_{bg}$ is created, and the node is labeled to indicate 
\react{5} must execute 50 times. \react{5}\ is not enabled in the specified 
initial state, so dependency edges are added to \react{3} and \react{8}, both 
of which generate RL and consequently enable \react{5}. Because \react{3} and \react{8} can both be executed 50 times from the initial state, the dependency graph is complete.
\begin{definition}[minimal leaf distance \mld]
	Define $\mld(\stChangeVec{i}) = 0$ if \stChangeVec{i}\ constitutes a leaf 
	node (i.e., 
	if \stChangeVec{i}\ is enabled in the model's initial state) and 
	$\mld(\stChangeVec{i}) 
	= \min(\{\ \mld(\stChangeVec{j})\ |\ j \in children(\react{i}) \}) + 1$ 
	in all 
	other cases.
\end{definition}

Intuitively, $\mld(\stChangeVec{i})$ represents the distance (in number of 
edges in the dependency graph) 
from node \stChangeVec{i}\ to a leaf node, or the number of unique transitions that
must execute to enable \stChangeVec{i}.
For example, in 
Figure~\ref{fig:depgraph}, 
$\mld({\react{5}}) = 1$.

\paragraph{Linear Algebra.}
Consider the set of rationals \rational. A \emph{linear vector space} in \rational, called $S$ is closed under linear combination,
i.e., $\vec{v}_i, \vec{v}_j \in S \implies \alpha_i\vec{v}_i + \alpha_j\vec{v}_j \in S$, where \ensuremath{\alpha_i, \alpha_j \in \rational}.
If a space $S_b$ with the same dimension is parallel to linear space $S$ but \emph{displaced}, \emph{offset}, or \emph{translated} by some vector $\vec{b}_b$, it is an \emph{affine space} to $S$. If $S_b$ is a valid affine space, there exists some mapping $\spaceMapping : S_b \mapsto S$, such that $\spaceMapping(\vec{v}) = \vec{v} - \vec{b}_b$ 
for any \ensuremath{\vec{v} \in S_b}.
The pseudo-``origin'' of this affine space is $\vec{b}_b$. In this paper, \emph{space} refers to both affine and linear spaces.
All $\rational^i, where \ i \in \nonNegInt$, are linear spaces,
\noindent and if space $S_b \subseteq S_a$, $S_b$ is a \emph{subspace} of $S_a$. Subspaces may be affine within the parent space, and inherit any affinity from that space. An example of an affine subspace of $\rational^3$ is any plane not intersecting the origin. 

\begin{definition}[generating set]\label{def:combinationSet}
    A \emph{generating set} 
    is a set of $q$ vectors whose linear combinations span a space of dimension $r$, thus $q \geqslant r$.
    If $q = r$, then it is a basis set for that space. 
    The vectors in this set are called \emph{generating vectors}.
\end{definition}

This work creates generating sets based on groups of transitions in a dependency graph, not guaranteed to be linearly independent or a basis set.
An affine space $S_b$ can be represented as $\{\ \vec{v}\ |\ \vec{v} = M_b\vec{x} + \vec{b}_b, \vec{x} \in \rational^q\ \}$, where $M_b$ is an $\speciesSetSize \times q$ matrix whose columns are vectors in any generating set spanning $S_b$. 

\begin{definition}[blanketing space]
The set of all transition vectors \vas in a VAS represents a generating set that, when displaced by the initial state, creates an affine space containing all reachable states. This space, called the \emph{blanketing space} \reachableSpace, is defined by $\{\ \vec{v}\ |\ \vec{v} = R\vec{a} + \vec{s}_0, \vec{a} \in \rational^\reactSetSize\ \}$.
\end{definition}

Because all states must have non-negative integer quantities of each species, they are first-orthant lattice points in $\reachableSpace$, but not all lattice points in $\reachableSpace$ are valid, reachable states, e.g., if a three-variable model ($\begin{bmatrix}\species{1} & \species{2} & \species{3} \end{bmatrix}^T$) has transitions $\begin{bmatrix}1 & 0 & 0\end{bmatrix}^T$ and $\begin{bmatrix}0 & -2 & 1\end{bmatrix}^T$, 
then no state can have $\species{3} > \initSt[2]$. 

If $\vec{s}$ is reachable, at least one solution of $\vec{s} = R\vec{a} + \vec{s}_0$ for $\vec{a}$, denoted as $\vec{a}_p$, satisfies $\vec{a}_{p} \in \nonNegInt^m$.
The elements of $\vec{a}_p$ correspond to the number of times each transition occurs to reach a state. More than one valid $\vec{a}_p$ may exist. 
Because not all transitions are enabled at every state (in a CRN, often due to a reaction consuming more of a species than a state can supply, placing a partial sum outside of the first orthant), not all permutations correspond to valid traces. All valid traces are thus constructed by a \emph{first orthant linear permutation} of all transitions in $\vec{a}_p$ to state $\vec{s}$. 
\noindent Also, the existence of an $\vec{a}_p$ does not imply corresponding traces, but the existence of a trace implies a corresponding $\vec{a}_p$.

Let  $\residmag_b$ represent the minimal distance from some vector $\vec{s}$ in vector space $V_s$ to affine subspace $S_b \subseteq V_s$ such that $\residmag_b = 0 \implies \vec{s} \in S_b$.
Given projection matrix $\projMat_b = M_b(M_b^TM_b)^{-1}M_b^T$, the formula $\residual_b = \projMat_b(\vec{s} - \vec{b}_b) - (\vec{s} - \vec{b}_b)$ gives the shortest \emph{residual vector}, $\residual_b$, such that $\vec{s} + \residual_b \in S_b$. Additionally, $\residmag_b = |\residual_b|$ is the shortest distance in a particular norm from $\vec{s}$ to space $S_b$. In the instance where $M_b^TM_b$ is singular, the Moore-Penrose pseudo-inverse\cite{penrose1955,dresdeninverse}, called $A^+$ for matrix $A$, is used and the formula becomes $\projMat_b = M_b(M_b^TM_b)^{+}M_b^T$. A pseudoinverse will be equal to an inverse if it exists.

\begin{definition}[solution space]\label{def:solSpace}
     For state formula \postUntil, the unique \emph{solution space} \solutionSpace is the smallest linear or affine space 
     such that 
    \noindent $\forall \vec{s}_s \in (\solutionSpace \cap \nonNegInt^m) \iff \vec{s}_s \models \postUntil$, where 
     $\solutionSpace = \{\ \vec{v}\ |\ \vec{v} = M_s\vec{x} + \vec{s}_p,\ \vec{x} \in \rational^m\ \}$. 
     In other words, \postUntil-formulas have solution spaces, if there exists $M_s$ and \state{p}, such that $\forall \vec{s}_s \models \postUntil \iff \exists \vec{x} \in \rational^m$ that satisfy $\vec{s}_s = M_s\vec{x} + \vec{s}_p$. That is, all valid states in \solutionSpace are solution states.
\end{definition}

In the motivating example, the solution space is a six-dimensional affine \emph{hyperplane} in the model's seven-dimensional space (i.e., $m=7$) created by restricting $G_{bg}$ (treated as a dimension) to $50$. 
Because explicit values of all species except $G_{bg}$ are arbitrary,
each of the six non-$G_{bg}$ species
trivially represent a unit-length basis vector in that species' dimension, creating an \emph{orthonormal basis set} (i.e., all basis vectors are unit-length and orthogonal to all others), as columns in matrix $M_s$. 
For a known solution $\vec{s}_p$,
all ``don't-cares'' may be chosen as $0$, i.e., $\vec{s}_p = \begin{bmatrix}
    0 & 0 & 0 & 0 & 0 & 50 & 0
\end{bmatrix}^T$. The solution space is simply this particular solution in addition to any linear combination of the basis, i.e., $\vec{s}_s = M_s\vec{x} + \vec{s}_p$, where $M_s$ contains the basis vectors as columns. Valid solution states restrict $\vec{s}_s \in \nonNegInt^{7}$, and the available transitions and choice of initial state $\vec{s}_0$ determines which of these solution states are \emph{reachable}, i.e., states which there exist traces to. All reachable states are contained within $\reachableSpace$ and therefore, $\reachableSpace \cap \solutionSpace = \emptyset$ implies that 
there are no reachable satisfying states. For VAS, the inverse is \emph{not} necessarily true, since $
\reachableSpace \cap \solutionSpace$ may exist, but not overlap any valid states. 
Because of this, constructing and reasoning about subspaces in VAS space is particularly useful for both time-bounded and unbounded eventually CSL properties, i.e., \cslEventually\ and \cslEventuallyUnbounded. In this paper, we restrict our focus to such properties, and contrive affine subspaces from update vectors which are used to reason about the existence of reachable states in the solution region. 


\section{Related Work}\label{sec:related}

\noindent \textit{Rare-event Simulation.} Using both the Monte Carlo method and stochastic simulation~\cite{Okamoto1959,Wald1945}, 
\emph{statistical model checking} (SMC) provides an \emph{estimate} of a property's probability.
Although some SMC tools (e.g., CRN-focused rare-event simulation tools~\cite{Gillespie2009,Daigle2011,Jegourel2012,Roh2010,Roh2011,Roh2016})  are equipped with rare event techniques (e.g., those surveyed in~\cite{Legay2016}), including importance sampling~\cite{Kahn1953}
and importance splitting~\cite{Kahn1951,Rosenbluth1955,Villen-Altamirano2011,Jegourel2013}, many require the user's input to provide an effective biasing scheme, and thus limited by user's deep insight of the model. 

\noindent \textit{Witness Trace Generation.} Witness trace (or counterexample) generation presented in~\cite{Han2007a} 
computes the most probable $k$-shortest paths to refute a probabilistic property for \emph{discrete-time Markov chains} (DTMCs). It was extended to generate approximate property-violation traces for CTMCs in~\cite{Han2007}, but was limited by scalability and did not produce a practical tool. Authors of~\cite{Aljazzar2009} presented a best-first search variant technique to scale up CTMC counterexample generation, 
but it cannot guarantee termination.
\noindent \emph{Bounded Model Checking} (BMC) was used to iteratively expand the partial state space of a CRN until its probability exceeds a given threshold~\cite{Ahmadi2024}. Although they approach the same problem as we do, their methods are orthogonal
due to their limit in the length of reachable traces and use of bounding constraints.
SAT-\cite{Wimmer2009} and SMT-based~\cite{Braitling2011} BMC was developed for DTMC counterexample generation, but could not be applied to CTMCs. \ragtimer~\cite{Israelsen2023,Taylor2023} leverages randomized compositional testing~\cite{Mcmillan2019} 
and the concurrency nature of CRNs to rapidly generate many probability-abstract traces to satisfying states, which are then used to construct a partial state graph to enable transient CTMC analysis.

Our proposed VAS-based technique
embeds known routes to satisfying states in closed subspaces, within which it can perform targeted state exploration. It outperforms \ragtimer (detailed in Section~\ref{sec:compareResults}) primarily because it is deterministic and produces no duplicate traces to satisfying states, while \ragtimer\ 
must filter out many duplicate traces \emph{after} random trace generation. 

\noindent \textit{VAS Reachability.}
Existing work focuses on the \emph{complexity analysis} of VAS reachability
%
~\cite{Czerwinski2020b,Leroux2011,Leroux2012VectorAS,Lipton1976,Rackoff1978}. 
Recently, Czerwinski~\cite{Czerwinski2021} and Leroux~\cite{Leroux2021} independently proved that VAS reachability was Ackermann-complete.
Separability between two VASes is analyzed in~\cite{Czerwinski2020}.
\noindent Our work 
seeks to find traces in two regions, namely, the reachable set and the satisfying set, which ought to be \emph{non-separable}.

\noindent \textit{Subspace Iteration.} A technique similar in name to the presented \isr method is \emph{subspace iteration}~\cite{stewartBook}, 
which is used to solve matrix-vector systems 
in steady-state PMC analysis.
In contrast, our methods \emph{create} a partial state space, represented as a rate matrix, but do not solve a matrix-vector system. The name similarity comes from the fact that they both iterate over vector subspaces.


\section{Search Algorithm}\label{sec:algorithm}

A significant challenge for CTMC transient analysis is restricting the size of the explicit state space while retaining useful information about the model. In the optimal case, CTMC transient analysis is performed on a partial state space containing only states with a non-zero probability of eventually reaching a satisfying state within a time bound. We propose two heuristics, \emph{Iterative Subspace Reduction} (\isr) and \textit{Single Distance Priority} (\sdp), which improve partial state-space construction by prioritizing a set of satisfying states along with states from which satisfying states are reachable in a limited number of steps.

Algorithm~\ref{alg:isrsearch} performs a priority-first search to construct a partial state space to be used for the transient analysis of PMC. This partial state space is constructed using either the \isr or \sdp heuristic. Both \isr and \sdp guarantee the inclusion of a specified number of satisfying states in the partial state space.

\begin{algorithm}[tbh]
\caption{\isr and \sdp search algorithm}\label{alg:isrsearch}
\begin{algorithmic}[1]
  \Input
  \Desc{$K$}{Number of desired satisfying states}
  \Desc{$\postUntil, \psi$}{State formula (\postUntil) in property $\psi$: \cslEventually\ or $\probOp_{=?}[\Diamond\ \postUntil ]$}
  \Desc{$(\vas, \initSt)$}{$m-$dimensional VAS model and initial state}
  \Desc{\texttt{method}}{Whether we are using \isr or \sdp}
  \EndInput
  \Output
  \Desc{\pMin}{Probability lower bound}
  \EndOutput
\State $pqueue \gets \{ \st_0 \}$
; $\foundSatStates \gets \emptyset$
; $\tranRateShort \gets \mathbf{0}$
; $\absSt \gets \vec{x} \in \nonNegInt^m\ s.t.\ \vec{x} \notin 
\reachableSpace \wedge \vec{x} \nvDash \postUntil$\label{algline:init}
\State{$Subspaces \gets \texttt{null},\ \stSet \gets \emptyset$}\label{algline:subspace}
\If{\texttt{method} $=$ \isr}\label{algline:isrFlag}
\State $D \gets \mathtt{setup\_dependency\_graph}(\vas)$
; $Subspaces \gets \mathtt{setup\_subspaces}(D)$\label{algline:setups}
\EndIf
\If{$dim(\solutionSpace \cap S_0 \cap \nonNegInt^m) = 1$}\label{algline:ensureTerminate}
    $K \gets 1$\Comment{To ensure termination};
\EndIf
\While{$pqueue \neq \emptyset \vee |\foundSatStates| < K$}\label{algline:mainLoop}
\State $\st \gets pqueue.\mathtt{dequeue}(\mathtt{method})$ \Comment{Using priority metric in Section~\ref{sec:priority}}
    \If{$\st \models \postUntil$}
        $\foundSatStates := \foundSatStates \cup \{ \st \}$; \textbf{continue}
    \EndIf
    \State $S_i \gets \mathtt{smallest\_space\_containing}(\st,\ Subspaces)$
    \For{$\stChangeVec{i},\ \propen{i}$ \textbf{in} $\mathtt{next}(\st)$}\label{algline:successors0}
        \If{\texttt{method} = \sdp $\wedge\ \st + \stChangeVec{i} \in S_i$}
            \State $\tranRateShort(\st, \st + \stChangeVec{i}) = \tranRateShort(\st, \st + \stChangeVec{i}) + \propen{i}(\st)$
            \If{$\st + \stChangeVec{i} \notin \stSet$}
                $pqueue.\mathtt{enqueue}(\st + \stChangeVec{i},\ \mathtt{method})$ 
            \EndIf
        \Else 
            \State 
            $\tranRateShort(\st, \absSt) = \tranRateShort(\st, \absSt) + \propen{i}(\st)$ 
        \EndIf
    \EndFor
\EndWhile
\While{$pqueue \neq \emptyset$}\Comment{Connect all unexplored states to $\absSt$};\label{algline:remainingStates}
    \State $\st \gets pqueue.\mathtt{dequeue}(\mathtt{method})$
    \If{$\st \models \postUntil$}
         $\foundSatStates := \foundSatStates \cup \{ \st \}$
    \EndIf
    \For{$\stChangeVec{i},\ \propen{i}$ \textbf{in} $\mathtt{next}(\st)$}\label{algline:successors1}
        \If{$\st + \stChangeVec{i} \in \stSet$}
        \State
            $\tranRateShort(\st, \st + \stChangeVec{i}) = \tranRateShort(\st, \st + \stChangeVec{i}) + \propen{i}(\st)$ 
            ; \textbf{continue}
        \EndIf
        \State $\tranRateShort(\st, \absSt) = \tranRateShort(\st, \absSt) + \propen{i}(\st)$ 
    \EndFor
\EndWhile
\State $\pMin \gets\ $\texttt{probabilistic\_model\_checking(state\_graph, $\psi$)}
\end{algorithmic}
\end{algorithm}

Inputs to Algorithm~\ref{alg:isrsearch} include $K$, the user-specified number of satisfying states to include in the partial state space; a state formula $\postUntil$ contained in the CSL property of interest $\psi$: \cslEventually\ or $\probOp_{=?}[\Diamond\ \postUntil ]$; an $m$-dimensional stochastic VAS model $\vas$ and its initial state $\initSt$; and \texttt{method}, the desired heuristic (\isr or \sdp). The output of Algorithm~\ref{alg:isrsearch} is \pMin, which is a lower bound on the CSL transient probability \cslEventually.

At a high level, Algorithm~\ref{alg:isrsearch} is a series of two loops (lines~\ref{algline:mainLoop} and~\ref{algline:remainingStates}). The first loop finds $K$ satisfying states, adding their successors to a priority queue based on the heuristic \texttt{method}. It then dequeues (explores) these successor states, constructing a partial state space that includes only states that eventually reach a satisfying state. When a satisfying state is found, it is added to the set \foundSatStates. Once the state space contains $K$ satisfying states, the second loop dequeues the remaining states, directing the remaining probability to an absorbing state \absSt. We assume $\absSt \not \models \postUntil$ to preserve a lower bound on the transient reachability probability. Finally, a probabilistic model checker (in our case, \texttt{stormpy}~\cite{Hensel2021}) evaluates the property $\psi$ for the partial state graph.

\subsection{Priority Metrics}

Our primary contribution is the development of \isr and \sdp. Both methods allow Algorithm~\ref{alg:isrsearch} to perform an on-the-fly priority-first search of an explicit state space for a stochastic VAS. 

\subsubsection{Single Distance Priority (\sdp)} Intuitively, \sdp prioritizes states with the shortest distance to the solution space \solutionSpace. Consider a projection matrix $\projMat_s$ of \solutionSpace for state \st. The state's distance $\residmag_s$ is defined by the magnitude of its residual to \solutionSpace (i.e., the difference between \st and its projection onto \solutionSpace). Specifically, $\residmag_s = |\projMat_s (\st - \st_p) - (\st - \st_p)|_1$, where $\st_p$ is an offset vector in \solutionSpace.

In the motivating example, $\postUntil$ is defined as $(G_{bg} = 50)$, constraining only species $G_{bg}$ and ignoring other species. For this model, \solutionSpace is a hyperplane orthogonal to dimension $G_{bg}$ at $G_{bg} = 50$. For an arbitrary state \st, $\residmag_s = |50 - \st[G_{bg}]|_1$. In other words, the distance is measured only in the dimension of interest ($G_{bg}$), and the distance is given by the absolute value of $50 - \st[G_{bg}]$. Intuitively, this represents the quantity of $G_{bg}$ that must be produced or consumed from the current state to reach a satisfying state. Consider
$\st_1[G_{bg}] = 29$ $\st_2[G_{bg}] = 36$. Because
$\residmag_{\st 1} = |50 - 29| = 21$ and $\residmag_{\st 2} = |50 - 36| = 14$, $\st_2$ is explored before $\st_1$.

\sdp requires that \solutionSpace can be represented as a closed subspace. That is, the state formula \postUntil must be a conjunction of substate formulas (i.e., \ensuremath{\postUntil = \bigwedge_{i=1}^{l} \postUntil_{i}}). 

For some species \species{j}, the substate formula $\postUntil_i$ ($i$ may differ from $j$) is given by $(\species{j} = \beta_i +\sum_{k = 1, k \neq j}^m \alpha_k \species{k})$, where $l \leqslant m$; $\alpha_k \in \mathbb{Q}_{\geqslant 0}$; and $\beta_i \in \nonNegInt$ represents the translation of \solutionSpace in the direction of $\species{j}$. Although $\beta_i \in \rational$ still retains subspace closure, $\beta_i$ is constrained to $\nonNegInt$ because valid VAS states must be non-negative lattice points. This ensures that at least one VAS state (specifically, $\species{j} = \beta_i$ with $\species{k} = 0, k \neq j$) satisfies $\postUntil_i$.

In addition, each species \species{k} contributes a basis vector $\vec{v}_k$ to the space of $\postUntil_i$. Specifically, $\vec{v}_k[j] = \alpha_k$, $\vec{v}_k[k] = 1$, and $\vec{v}_k[w] = 0$ for $w \neq j,k$. If $\alpha_k = 0$ (i.e., \species{k} is not present in the substate formula), $\vec{v}_k$ is a unit basis vector in the direction of \species{k}. Thus, $\postUntil_i$ includes $m-1$ basis vectors and forms a hyperplane in $\rational^m$. This subspace is offset by the vector $\vec{b}_i$, a vector with $\vec{b}_i[j] = \beta_i$ and $\vec{b}_i[w] = 0$ for $w \neq j$.
Thus, $\postUntil_1,\ \postUntil_2,\ \cdots, \postUntil_l$ each contribute to subspaces whose intersections comprise \solutionSpace. The intersection of these spaces can be used to find a basis and a particular solution to \solutionSpace.

The subspace encoded in $\postUntil_i$ also implies a \emph{normal vector} $\vec{n}_i$, which is orthogonal to all vectors in $\postUntil_i$. The normal vector contains elements $\vec{n}_i[j] = 1$, $\vec{n}_i[k] = -\alpha_k$, and $\vec{n}_i[j] = 0$ for $w \neq j,k$. Because each space is a hyperplane, normal vectors can be compared to reveal redundant or contradictory subformulas. Subformulas with parallel normal vectors are redundant (i.e., the spaces encode identical information) or contradictory (i.e., no state is contained in their intersection). Although this is not necessary, the Gram-Schmidt process~\cite{Leon2013} can be used to produce an orthogonal basis set.

The existence of a valid solution space depends on whether there is an intersection of the substate formulas' spaces. To find the intersection between two spaces with $ p \leqslant m $ and $ q \leqslant m $ respective generating vectors, define $S_a = \{\ M_a \vec{x}_a + \vec{b}_a\ |\ \vec{x}_a \in \rational^p\ \}$ and $S_b = \{M_b \vec{x}_b + \vec{b}_b\ |\ \vec{x}_b \in \rational^q\ \}$ such that $S_a = S_b$. Therefore, $M_a \vec{x}_a + \vec{b}_a = M_b \vec{x}_b + \vec{b}_b \Leftrightarrow M_a\vec{x}_a - M_b\vec{x}_b = \vec{b}_b - \vec{b}_a$. Thus, $\begin{bmatrix} M_a & - M_b  \end{bmatrix} \begin{bmatrix} \vec{x}_a & \vec{x}_b  \end{bmatrix}^T = \vec{b}_b - \vec{b}_a$ or $\mathbf{A}\hat{x} = \vec{b}_b - \vec{b}_a$ may be solved for $\hat{x} = \begin{bmatrix} \vec{x}_a & \vec{x}_b  \end{bmatrix}^T$, where $\mathbf{A} = \begin{bmatrix} M_a & - M_b  \end{bmatrix}$. The \emph{nullspace} $\mathcal{N}(\mathbf{A}) = \{\ \vec{x}_n\ |\ \mathbf{A}\vec{x}_n = \vec{0}\ \}$ shares all basis vectors for this intersection, and a particular solution/translation vector may be easily obtained once the system is in row-echelon form. 

For example, consider a model with species \species{1}, \species{2}, and \species{3}. Under assumptions made by \sdp, the following substate formulas are legal: $\postUntil_1: \species{1} = \species{2} + \beta_1$ and $\postUntil_2: \species{1} = 2 \species{2}  + \species{3}$. Thus, $\postUntil$ is $( \species{1} = \species{2} + \beta_1) \land (\species{1} = 2 \species{2}  + \species{3})$. 
$\postUntil_1$ corresponds to the normal vector $\begin{bmatrix} 1 & -1 & 0 \end{bmatrix}^T$, and $\postUntil_2$ corresponds to the normal vector $\begin{bmatrix} 1 & -2 & -1 \end{bmatrix}^T$. Offset vectors are $\begin{bmatrix} \beta_1 & 0 & 0 \end{bmatrix}^T$ (offset of $\beta_1$ in the direction of \species{1}) and $\vec{0}$ (no offset in the direction of \species{1}). 
\noindent Basis vectors for $\postUntil_1$ are $\begin{bmatrix} \alpha_3 & 0 & 1\end{bmatrix}^T = \begin{bmatrix} 0 & 0 & 1\end{bmatrix}^T$ and $\begin{bmatrix} \alpha_2 & 1 & 0\end{bmatrix}^T = \begin{bmatrix} 1 & 1 & 0\end{bmatrix}^T$. 
Similarly, basis vectors for $\postUntil_2$ are $\begin{bmatrix} 2 & 1 & 0\end{bmatrix}^T$ and  $\begin{bmatrix} 1 & 0 & 1\end{bmatrix}^T$. Since these two (hyper)planes are not parallel (their normal vectors differ), an intersection exists, and its location is a function of $\beta_1$.

Disjoint substate formulas may break subspace closure, so it is currently not considered, although future work will explore multiple solution subspaces and $\postUntil-$formula reduction by which of these intersect with $\reachableSpace$. In this case, DeMorgan's Law does not apply, as $\neg(\postUntil_i \vee \postUntil_j) = \neg \postUntil_i \wedge \neg \postUntil_j$. If $\postUntil_i$ corresponds to space $S_{si}$, then $\neg \postUntil_i$ corresponds to $S_{si}^C$, which is not a closed subspace. However, our existing requirements are more generalizable than those in \ragtimer~\cite{Israelsen2023,Taylor2023}.

\subsubsection{Iterative Subspace Reduction (\isr)}\label{sec:subcons}
Intuitively, \isr creates a set of nested closed subspaces through which the satisfying traces with the highest probabilities are likely to travel. If some transition \stChangeVec{B} depends on the execution of \stChangeVec{A} to become enabled, then any trace in which \stChangeVec{A} always executes before \stChangeVec{B} must pass through a space with generating set $\{\ \stChangeVec{A},\ \stChangeVec{B}\ \}$, followed by a space with basis vector $\{\ \stChangeVec{B}\ \}$. We impose an ordered index on these subspaces (i.e., $S_0,S_1,\cdots,S_\subspacesCount$) such that the following conditions hold:  
(1)~$j > i \implies S_j \subseteq S_i$, and
(2)~$S_0 \cap \solutionSpace \neq \emptyset \implies \solutionSpace \cap S_\subspacesCount \neq \emptyset$. 
Then, a subspace $S\subspacesCount$ is constructed to overlap the satisfying space \solutionSpace. $\isr$ seeks satisfying states in $S\subspacesCount$. In Line~\ref{algline:setups} of Algorithm~\ref{alg:isrsearch}, the function $\mathtt{setup\_subspaces}(D)$ produces this list of nested subspaces, described more formally in the following definitions and theorems.

\begin{definition}[indexed subspace]\label{def:indexedsub}
Let affine subspace $S_i$ assigned \emph{index} $i \in \nonNegInt^{\leqslant \subspacesCount}$ be $S_i \triangleq \{\ M_i\vec{x} + \st_0 + \vec{f}\ |\ \vec{x} \in \rational^m \}$, where columns of $M_i$ are $M_i \triangleq \begin{bmatrix}\cdots, \stChangeVec{j}, \cdots \end{bmatrix}$ and $j \in J$ such that $J \triangleq \{\ k\ |\ \mld(\react{k}) \geqslant i\ \}$. $\vec{f}$ is an additional displacement vector such that $\vec{f} \in \{\ M_0\vec{x} + \st_0\ |\ \vec{x} \in \rational^m \}$ and $\initSt + \vec{f} \in \solutionSpace$.
\end{definition}

More details on the choice of $\vec{f}$ are provided in Appendix~\ref{apx:offvec}. The index $i$ of $S_i$ is the smallest \mld in the dependency graph such that any update vector with that \mld is permitted in the subspace's generating set. The generating vectors for indexed subspaces are required to be update vectors in the dependency graph, and all indexed subspaces are contained within the blanketing spaces $\reachableSpace$ of the VAS.
Not all vectors in $ S_i $ are valid or reachable. However, any state reachable from another state in $ S_i $ through the transitions $ \stChangeVec{j} $ (where $ j \in J $ and $ \mld(\stChangeVec{j}) \geqslant i $) is guaranteed to be a state within $ S_i $. 
The motivating example yields two indexed subspaces: $S_0$ and $S_1$. $S_1$ has only \stChangeVec{5} in its generating set, and the generating set for $S_0$ is $\{ \stChangeVec{3}, \stChangeVec{8}, \stChangeVec{5} \}$ --- every transition in the dependency graph.


Theorems~\ref{thm:ineqempty} and~\ref{thr:sat} describe guarantees regarding the feasibility of \isr. 

\begin{theorem}\label{thm:ineqempty}
    If there is at least one reachable solution state, $S_I \neq \emptyset$.
\end{theorem}
\begin{proof}
    If any solution state is reachable, \cite{Israelsen2023} proves that at least one solution state is reachable using only transitions in the dependency graph.
    By construction, $S_0$ contains all states reachable by those transitions.
    Any solution state is in $\solutionSpace$, thus $S_I \neq \emptyset$. \qed
\end{proof}
\begin{remark}\label{remark:nest}
    For \emph{indexed} subspaces $S_i$ and $S_j$, $j > i \implies S_j \subseteq S_i$. By definition, $S_i$ and $S_j$ share an offset vector, and $j > i$ implies that the generating set of $S_j$ is a subset of the generating set of $S_i$. Consider $S_i$ as described in Definition~\ref{def:indexedsub}. If $ S_0 \cap \solutionSpace \neq \emptyset $, it follows that $ S_i \cap \solutionSpace \neq \emptyset $ for all $ j > i $, and $ S_j \subseteq S_i $. Additionally, since $ \st_0 \in S_i^* $, all states reachable in $ S_i $ can be reached from the generating set of $ S_0 $. The construction of these subspaces occurs only if $ S_0 \cap \solutionSpace \neq \emptyset $, as a valid dependency graph is required a priori.
\end{remark}

\begin{theorem}\label{thr:sat}
    For a set of indexed subspaces and \solutionSpace, $S_i \cap \solutionSpace \neq \emptyset$.
\end{theorem}
\begin{proof}
    By construction, $\initSt + \vec{f} \in \solutionSpace$, and $\initSt + \vec{f} \in S_i$ for all $i$. \qed
\end{proof}
\begin{remark}
    Although $\solutionSpace \cap S_i \neq \emptyset$ is not guaranteed to include valid states, Algorithm~\ref{alg:isrsearch} only explores valid states and thus explores states closer to each $S_i$. 
\end{remark}

\subsection{Comparing Prioritization in \isr and \sdp}\label{sec:priority}

Algorithm~\ref{alg:isrsearch} depends on a heuristic to determine what state to dequeue first. Thus, \isr and \sdp both impose a partial order on queued states to create this heuristic. \sdp is a simple approach: it chooses the state with the shortest 1-norm distance to \solutionSpace. In contrast, \isr compares $\st_i \in S_p$ and $\st_j \in S_q$ by calculating $\max(p)$ and $\max(q)$, prioritizing $\st_i$ if $p > q$ (i.e., $S_p \subseteq S_q$), or $\st_j$ if $p > q$. If $p = q$, the state with the smaller residual is prioritized. Because $p > q \iff S_p \subseteq S_q \iff (\residmag_p = 0 \implies \residmag_q = 0)$, distance computations start at $S_\subspacesCount$ and short-circuit at the first index where zero-distance is found.


\begin{theorem}\label{thr:termination} 
    Algorithm~\ref{alg:isrsearch} terminates for both \isr and \sdp when $\postUntil = \bigwedge \postUntil_i$ and $\postUntil_i: \species{j} = \beta_i$. Because \isr requires this condition, it always terminates.
\end{theorem}
\begin{proof}
    If there exists exactly one solution state, then $dim(\solutionSpace) = 0$. Algorithm~\ref{alg:isrsearch} checks for this condition on Line~\ref{algline:ensureTerminate} and assigns $K=1$. 
    Because \postUntil\ is composed of a conjunction of substate formulas $\postUntil_i$, where $\postUntil_i$ is $\species{j} = \beta_i$, $dim(\solutionSpace) > 0$ implies that $\solutionSpace \cap \nonNegInt^m \neq \emptyset$.
    The number of valid solution states $|\solutionSpace \cap \nonNegInt^m|$ is infinite, as the spaces of each $\postUntil_i$ is orthogonal to the $\species{j}$-axis, so $\forall j. \postUntil \perp \species{j}$. 
    Since $\beta_i \in \nonNegInt$, for any $\st_\postUntil \in \solutionSpace \cap \nonNegInt^m$, a new state $\st_\postUntil'$ may be constructed to be equivalent to $\st_\postUntil$ on all axes except $\st_\postUntil'[p] = \st_\postUntil[p] + c,\ c \in \nonNegInt$ for some $p \neq j$.
    Termination thus depends on the existence of $K$ solution states and their discoverability by both \isr and \sdp. 
    
    Because \sdp is an exhaustive priority-first search of a state space, if $K$ satisfying states exist, \sdp will eventually find each of them and termination is vacuously true.
    In contrast, \isr explores only successors available via the transitions in a dependency graph. Because a dependency graph is cycle-free and encodes exactly the shortest traces, any trace in the VAS state graph corresponding to a path from root to leaf nodes in the dependency graph (i.e., composed of only transitions specified in the dependency graph and with a VAS transition count greater than or equal to the sum of dependency graph edge weights) leads to a \emph{unique} satisfying state. 
    Thus, if $K$ satisfying states exist, \isr will eventually terminate after finding them. If \isr explores all satisfying states in $S_0$, it has found all possible traces and will terminate. Because \isr greedily expands successors (i.e., only explores those which move it closer to the solution), 
    it may terminate \emph{before} $K$ solution states are reached.
    \qed
\end{proof}
\begin{remark}
    In Theorem~\ref{thr:termination}, termination requires for \solutionSpace that a subset of axes be restricted to a constant positive integer value. This is stricter than the general formula for \solutionSpace described under assumptions for \sdp. To construct a dependency graph (a requirement for \isr), the stricter requirements of Theorem~\ref{thr:termination} are necessary. In \sdp, the requirements can be loosened, but termination is no longer guaranteed. All realistic case studies we have encountered fit the requirements of the stricter theorem.
\end{remark}

\section{Implementation}\label{sec:implementation}
Both \isr and \sdp were implemented in Python using \texttt{numpy}\cite{harris2020array}, as part of a prototype tool, Wayfarer,
which interfaces with the \storm Python API~\cite{Hensel2021} by constructing a sparse matrix and labeling and then passing that matrix to \texttt{stormpy} for PMC.
A prototype version of this tool is publicly available
\footnote{\url{https://github.com/fluentverification/wayfarer}}. 
It rapidly finds a partial state space that generally contains 
\emph{$\min(K, |\{ \st_s\ |\ \st_s \in S_0 \cap \solutionSpace \cap \nonNegInt^m \}|)$ unique satisfying states}, where $K$ is a user-specified parameter. It currently supports CRN models in the \ragtimer~\cite{Taylor2023,Israelsen2023} input format, which is simply a plain text file that provides information about reactants, products, and reaction rate constants, and the satisfying state formula of a CRN. Because all tested models have more than one valid satisfying state, the check on line~\ref{algline:ensureTerminate} of Algorithm~\ref{alg:isrsearch} was not included in the prototype.

\paragraph{Floating Point Imprecision.}
In early tests, floating-point imprecision affected the order of exploration. 
If two states have zero L2 distance $S_i$, floating-point calculations of those distances
may return extremely low values, such as $10^{-99}$ or $10^{-100}$.
These errors effectually prioritize for the state with the slightly smaller error, affecting the exploration order.
Since spaces in our VAS need only cover lattice points,
distances calculation uses rational numbers and an L1 norm. Transition rates, and thus precision in PMC, are still stored as floating-point.

\subsubsection{Limitations.}
One limitation is all of the solution points must exist within a closed subspace or a single point.
This is possible when one or more species are marked as ``don't care'', such as in the motivating example, where all species except $G_{bg}$ are marked as such. Here, we can define \solutionSpace's basis set as, for each don't-care species, the unit vector in that direction, compiled into matrix $M_s$, offset by a particular solution $\st_p$, chosen by setting all ``don't-cares'' to zero.

Although in theory this limits state formulas to exclude inequalities such as $>, <, \leqslant,$ and $\geqslant$, often in practice a closed ``boundary condition'' can be created for inequality formulas.
Additionally, the input CRN format --- the same as for \ragtimer~\cite{Israelsen2023,Taylor2023} --- currently only supports strict equality or ``don't-cares'' for state formulas and unit-step increment and decrement in update vectors.

\paragraph{Upper Bound} Creating an upper bound on probability with these methods is nontrivial. Although it is possible to [TODO]
\section{Results and Discussion}\label{sec:results}
All results were obtained on an AMD Ryzen Threadripper 12-Core 3.5 GHz Processor and 132 GB of RAM, running Ubuntu 22.04 LTS. We allocated one CPU and
unlimited RAM (though our tool needed very little)
to ten challenging case studies and compared our 
method's results to those 
of other probabilistic verification tools.
Due to the small, targeted state spaces generated,
the vast majority of runtime is spent on state expansion, with
negligible time on PMC. 

\subsubsection{Comparison to the State-of-the-Art Tools}~\label{sec:compareResults}
Of particular interest in comparison is the recent tool \ragtimer~\cite{Israelsen2023}, a recent tool that uses assume-guarantee reasoning to non-deterministically generate satisfying traces. We also compared our methods against \stamina~\cite{Jeppson2023}, a state space truncation tool for large and infinite models, and \storm~\cite{Hensel2022}, a high-performance probabilistic model checker. 

Our test suite involved 10 CTMC models, including our motivating example, from the \prism\ benchmark suite~\cite{Kwiatkowsa2012} and QComp~\cite{Qcomp2020}, and the biochemical models originally used to test \ragtimer~\cite{Israelsen2023}. Models, described in Appendix~\ref{apx:models}, were chosen based on whether they could be encoded as VAS, and if they had properties of the form \cslEventually. For multiple models, \storm was unable to return a result due to out-of-memory (OOM) or timeout, and \ragtimer was incompatible with several models due to custom transition rate functions. Both \isr and \sdp are targeted at rare events (highlighted), which have probability $\leqslant 10^{-3}$. On non-rare events, these methods are not expected to perform as well.

\begin{table}[htbp]
\centering
\resizebox{\textwidth}{!}{%
\begin{tabular}{C{1.3cm} C{1.8cm} C{0.7cm} C{2.0cm} C{1.0cm} C{2.0cm} C{1.0cm} C{2.0cm} C{2.0cm} C{2.0cm} C{2.0cm}}
\toprule
\textbf{Model} & \textbf{Property} 
& \multicolumn{2}{c}{\textbf{Wayfarer \sdp}} 
& \multicolumn{2}{c}{\textbf{Wayfarer \isr}} 
& \multicolumn{2}{c}{\textbf{\ragtimer}} 
& \multicolumn{2}{c}{\textbf{\stamina}} 
& \textbf{\storm} \\
& & K & \pMin & K & \pMin & n & \pMin & \pMin & \pMax & \probOp \\
\midrule
\multirow{3}{*}{CM}
& p\_CM1          & 1       & 1            & 1       & 1            & – & –            & 1            & 1            & 1             \\
& p\_CM2          & $10^{4}$ & $5.20\textsc{e}{-4}$  & $10^{4}$ & $1.38\textsc{e}{-2}$  & – & –            & 0.01376307  & 0.013764839 & 0.013763076  \\
\rowcolor{gray!20}
& p\_CM3          & 100     & $1.73\textsc{e}{-22}$ & $10^{4}$ & $1.76\textsc{e}{-18}$ & – & –            & 0            & $1.77\textsc{e}{-6}$  & $2.61\textsc{e}{-17}$ \\
\midrule
\rowcolor{gray!20}
SSPD 
& p\_SSPD        & 10    & $2.99\textsc{e}{-7}$  & 10    & $1.94\textsc{e}{-16}$ & $10^{3}$ & $1.40\textsc{e}{-7}$ & 0            & 0.02504      & $2.99\textsc{e}{-7}$  \\
\midrule
\rowcolor{gray!20}
EFC  
& p\_EFC         & 1       & $1.53\textsc{e}{-7}$  & 1       & $2.02\textsc{e}{-78}$ & $10^{4}$ & $4.38\textsc{e}{-10}$& 0            & 0.02504      & $1.74\textsc{e}{-7}$  \\
\midrule
\rowcolor{gray!20}
MYP  
& p\_MYP         & $10^{3}$ & $5.49\textsc{e}{-76}$ & 100     & $3.77\textsc{e}{-16}$ & 100       & $5.52\textsc{e}{-27}$ & $1.43\textsc{e}{-6}$ & $1.49\textsc{e}{-5}$ & \texttt{TIMEOUT} \\
\midrule
\rowcolor{gray!20}
SMR  
& p\_SMR         & 10    & $6.71\textsc{e}{-12}$ & 10    & $7.51\textsc{e}{-14}$ & 1         & $1.43\textsc{e}{-15}$ & $2.48\textsc{e}{-7}$ & $2.54\textsc{e}{-7}$ & $2.48\textsc{e}{-7}$  \\
\midrule
\rowcolor{gray!20}
\multirow{2}{*}{P53}

& p\_P53\_1      & 10    & $4.35\textsc{e}{-24}$ & 10    & $4.46\textsc{e}{-32}$ & –         & –            & 0            & 0.304        & \texttt{OOM} \\
\rowcolor{gray!20}
& p\_P53\_2      & $10^{3}$ & $2.38\textsc{e}{-50}$ & $10^{3}$ & $5.73\textsc{e}{-66}$ & –         & –            & 0            & 0.997        & \texttt{OOM} \\
\midrule
MGD  
& p\_MGD         & 1       & $1.22\textsc{e}{-146}$& 1       & $1.22\textsc{e}{-146}$& –         & –            & 0.0537       & 0.0545       & 0.0543         \\
\midrule
SID  
& p\_SID         & $10^{4}$ & $1.04\textsc{e}{-172}$& $10^{5}$ & $1.04\textsc{e}{-172}$& –         & –            & 0.0422       & 0.0422       & 0.0423         \\
\midrule
\multirow{2}{*}{HT}
& p\_HT1         & 10       & $4.63\textsc{e}{-88}$ & 1       & $2.79\textsc{e}{-108}$ & –         & –            & 0.00030      & 0.00030      & \texttt{TIMEOUT} \\
& p\_HT2         & 1       & $1.50\textsc{e}{-46}$ & 1       & $8.54\textsc{e}{-48}$ & –         & –            & 0.00030      & 0.00030      & \texttt{TIMEOUT} \\
\midrule
\rowcolor{gray!20}
TQN  
& p\_TQN         & 10      & $3.24\textsc{e}{-44}$ & 1       & $4.44\textsc{e}{-49}$ & –         & –            & $5.78\textsc{e}{-7}$ & $2.23\textsc{e}{-6}$ & $5.78\textsc{e}{-7}$ \\
\bottomrule
\end{tabular}%
}
\caption{Probabilistic results for \sdp, \isr, \ragtimer, \stamina\ and \storm. Rare events are highlighted. A dash indicates the model was unsupported.}
\label{tab:verification-finally}
\end{table}

Tables~\ref{tab:verification-finally} and~\ref{tab:tool-benchmark-table} present results for \isr and \sdp, and compare them against \ragtimer, \stamina, and as a control, the probabilistic model checker \storm. Due to limitations in tool compatibility, \ragtimer could not be run on most models. On some large and unbounded models, \storm was unable to generate a result due to timeout or memory out. In particular, \ragtimer was unable to build traces or compute probabilities for the \emph{cancer metabolism model} (CM), as \ragtimer's algorithm requires the traversal of the model's very large dependency graph, exceeding reasonable recursion depth limits. While \stamina generally produced superior probabilistic results, it did so at a significantly higher runtime, highlighting the efficiency of Wayfarer. In Table~\ref{tab:verification-finally}, $K$ is the number of requested satisfying states, and for \ragtimer, $n$ is the number of satisfying traces. In both cases, the presented $K$ or $n$ is the highest value before either tool was unable to immediately substantially improve on their results. Table~\ref{tab:tool-benchmark-table} omits state space size ($N$) and memory usage for \ragtimer, as it generates potentially duplicated \emph{traces} (which often contain the same state multiple times), and utilizes the hard disk for data, as opposed to exclusively RAM.

Wayfarer, regardless of whether \isr or \sdp is used, is able to \emph{rapidly} generate a partial state graph that appears to contain all or most \emph{shortest} traces. On models with more complex dependency graphs, such as the MYP, \isr outperforms \sdp in runtime. However, in the MYP, it is unable to close the gap to the actual probability of around $10^{-6}$. We suspect this is due to the high-probability cycle in that state graph identified by~\cite{Taylor2023}, which, as it contains ``irrelevant'' transitions (those not included in shortest traces) is ignored by \isr. It is likely this limitation can be mitigated by a cycle-and-commute extension similar to~\cite{Taylor2023} or some kind of abstraction. In fact, as \isr restricts transitions to those in the dependency graph, the state space it generates is well-suited to such methods. On models with simpler dependency graphs, such as the SSPD and EFC models, \sdp obtained more accurate values of \pMin than \isr. On the EFC model, \sdp achieved a \pMin very close to the actual probability at about $1/3$ of the state space used by \storm. Wayfarer, utilizing \isr and \sdp, consistently delivers rapid results, \emph{achieving partial state graphs in under 2 minutes for large models, a stark contrast to the hours required by competing tools.} For instance, in the MYP model, \isr produced results in under $30s$, while \stamina required nearly 4 hours, and \storm timed out. Even \ragtimer, the current state-of-the-art for rare events, took longer and yielded a lower \pMin. Since both \isr and \sdp terminate exceptionally quickly, it becomes trivial for users to try both methods if it is unknown which is better suited to a model. This rapid analysis capability allows researchers to iterate more quickly on model design and testing.
\begin{table}[htbp]
\centering
\resizebox{\textwidth}{!}{%
\begin{tabular}{C{1.5cm} C{1.8cm} 
                C{1.6cm} C{1.6cm} C{1.6cm} 
                C{1.6cm} C{1.6cm} C{1.6cm}
                C{1.6cm}
                C{1.6cm} C{1.6cm} C{1.6cm}
                C{1.6cm} C{1.6cm} C{1.6cm}}
\toprule
\textbf{Model} & \textbf{Property} 
& \multicolumn{3}{c}{\textbf{Wayfarer \sdp}} 
& \multicolumn{3}{c}{\textbf{Wayfarer \isr}} 
& \textbf{\ragtimer} 
& \multicolumn{3}{c}{\textbf{\stamina}} 
& \multicolumn{3}{c}{\textbf{\storm}} \\
& & \(\mathbf{t}\) & \stSetSize & mem
& \(\mathbf{t}\) & \stSetSize & mem 
& \(\mathbf{t}\)
& \(\mathbf{t}\) & \stSetSize & mem 
& \(\mathbf{t}\) & \stSetSize & mem\\
\midrule
\multirow{3}{*}{CM}
& p\_CM1 
  & 0.001   & 2      & 68.27MB 
  & 0.001   & 2      & 68.99MB 
  & –       & 0.572   & 2      & 80.08MB 
  & 0.028   & 1      & 69.94MB \\
& p\_CM2 
  & 40.973  & 64346  & 348.94MB 
  & 705.403 & 352173 & 1.87GB  
  & –       & 55.189  & 194376 & 212.15MB 
  & 59.353  & 2251898& 1.26GB  \\
\rowcolor{gray!20}
& p\_CM3 
  & 0.884   & 1665   & 74.42MB 
  & 1045.415& 532281 & 2.86GB  
  & –       & 56.252  & 194167 & 212.50MB 
  & 57.321  & 2176403& 1.24GB  \\
\midrule
\rowcolor{gray!20}
SSPD 
& p\_SSPD 
  & 0.019   & 83     & 68.63MB 
  & 0.008   & 42     & 70.02MB 
  & 63.51   & 0.763   & 53     & 68.41MB 
  & 0.023   & 81     & 60.94MB \\
\midrule
\rowcolor{gray!20}
EFC  
& p\_EFC  
  & 0.038   & 102    & 69.00MB 
  & 0.024   & 52     & 69.96MB 
  & 735.36  & 1.151   & 89     & 72.57MB 
  & 0.038   & 298    & 64.12MB \\
\midrule
\rowcolor{gray!20}
MYP  
& p\_MYP  
  & 61.286  & 3744   & 81.37MB 
  & 26.307  & 42165  & 214.16MB 
  & 66.75   & 13969.482 & 192496 & 205.64MB 
  & \multicolumn{3}{c}{TIMEOUT} \\
\midrule
\rowcolor{gray!20}
SMR  
& p\_SMR  
  & 0.231   & 522    & 70.30MB 
  & 0.021   & 41     & 70.30MB 
  & 23.06   & 27.552  & 67210  & 111.77MB 
  & 4.716   & 277830 & 206.89MB \\
\midrule

\rowcolor{gray!20}
\multirow{2}{*}{P53}

& p\_P53\_1
  & 0.049   & 233    & 69.87MB 
  & 0.011   & 52     & 70.25MB 
  & –       & 159.919 & 178071 & 175.71MB 
  & \multicolumn{3}{c}{OOM}\\
\rowcolor{gray!20}
& p\_P53\_2
  & 0.736   & 2414   & 74.80MB 
  & 0.018   & 62     & 69.93MB 
  & –       & 2070.626& 178071 & 174.12MB 
  & \multicolumn{3}{c}{OOM} \\
\midrule
MGD  
& p\_MGD  
  & 0.660   & 1261   & 72.38MB 
  & 0.655   & 1261   & 72.55MB 
  & –       & 227.058 & 57846  & 130.00MB 
  & 19.604  & 191250 & 177.25MB \\
\midrule
SID  
& p\_SID  
  & 68.793  & 95341  & 416.87MB 
  & 70.717  & 95341  & 416.97MB 
  & –       & 116.039 & 273720 & 357.62MB 
  & 73.367  & 737280 & 653.89MB \\
\midrule
\multirow{2}{*}{HT} 
& p\_HT1  
  & 0.011   & 62    & 69.00MB 
  & 0.081   & 13     & 69.56MB 
  & –       & 0.946   & 111    & 66.43MB 
  & \multicolumn{3}{c}{TIMEOUT}\\
& p\_HT2  
  & 0.006   & 28     & 68.50MB 
  & 0.005   & 22     & 69.72MB 
  & –       & 0.956   & 111    & 66.45MB 
  & \multicolumn{3}{c}{TIMEOUT} \\
\midrule
\rowcolor{gray!20}
TQN  
& p\_TQN  
  & 0.365   & 1562   & 72.94MB 
  & 0.109   & 513    & 71.31MB 
  & –       & 1.969   & 5467   & 76.99MB 
  & 4.461   & 523776 & 207.24MB \\
\bottomrule
\end{tabular}%
}
\caption{Resources used by each tool. \(t\) represents the runtime (in seconds); \stSetSize\ represents the size of the state space; and \emph{mem} represents the memory usage. \stSetSize\ and memory usage are omitted for \ragtimer as it may duplicate states and utilizes the hard drive.}
\label{tab:tool-benchmark-table}
\end{table}

On non-rare events, as anticipated, \isr and \sdp performed less effectively than \stamina. However, in all rare events except the TQN, both methods demonstrated significant performance improvements, particularly in runtime, over all tools, especially \ragtimer. Notably, the high memory usage observed in Wayfarer can be attributed to its proof-of-concept Python implementation; we anticipate that an optimized version in C++ or Rust would exhibit memory usage proportional to its compact state spaces.

\noindent\textit{Rare-event simulation.}
The statistical model checking tool \modes\ in the \modestToolset\ 
computed rare-event probability estimates efficiently for our motivating example, with estimates from \modes\ closely matching the
the probability reported in~\cite{Donovan2013}.
However, because \modes\ requires a compositional importance function
for rare-event simulation, it prevents global variables and 
requires an in-depth understanding of a CRN model and the 
\modest language.




\section{Conclusion}
We present novel heuristics, \isr\ and \sdp, to find traces to rare events in stochastic VAS.
Both \isr and \sdp scale well on models with small dependency graphs (even with large state spaces). \sdp is applicable to a broader set of models, including those where a dependency graph cannot be constructed. 
Benchmarking results show that our methods outperformed many existing tools for challenging rare-event CRN case studies.
Since the paths \emph{excluded} from the reported probabilities here are those with cycles and ``irrelevant'' transitions ($\{ \stChangeVec{i}\ |\ \stChangeVec{i} \perp S_0\ \}$), as our future work, we plan to collapse all states outside of $S_0$ onto points in $S_0$, creating an abstracted CTMC based on the partial state space found by \isr to enable transient analysis. \sdp and \isr both perform extremely well on rare-events that challenge state of the art tools. Though our methods are formulated for and tested on VAS, we suspect they also apply to VAS over rationals.


\newpage
\bibliography{references/refs,references/myPubs,references/trie}

\begin{thebibliography}{10}

\bibitem{Ahmadi2024}
Mohammad Ahmadi, Lukas Buecherl, Chris~J. Myers, Zhen Zhang, Chris Winstead, and Hao Zheng.
\newblock Rare-event guided analysis of infinite-state chemical reaction networks.
\newblock In Jane Hillston, Sadegh Soudjani, and Masaki Waga, editors, {\em Quantitative Evaluation of Systems and Formal Modeling and Analysis of Timed Systems}, pages 196--212, Cham, 2024. Springer Nature Switzerland.

\bibitem{Aljazzar2009}
Husain Aljazzar and Stefan Leue.
\newblock Directed explicit state-space search in the generation of counterexamples for stochastic model checking.
\newblock {\em IEEE Transactions on Software Engineering}, 36(1):37--60, 2009.

\bibitem{Aziz2000}
Adnan Aziz, Kumud Sanwal, Vigyan Singhal, and Robert Brayton.
\newblock Model-checking continuous-time markov chains.
\newblock {\em ACM Trans. Comput. Logic}, 1(1):162--170, July 2000.

\bibitem{Braitling2011}
Bettina Braitling, Ralf Wimmer, Bernd Becker, Nils Jansen, and Erika {\'A}brah{\'a}m.
\newblock Counterexample generation for markov chains using smt-based bounded model checking.
\newblock In {\em Formal Techniques for Distributed Systems}, pages 75--89. Springer, 2011.

\bibitem{Budde2019}
Carlos~E. Budde, Pedro~R. D'Argenio, and Arnd Hartmanns.
\newblock Automated compositional importance splitting.
\newblock {\em Science of Computer Programming}, 174:90--108, 2019.
\newblock URL: \url{https://www.sciencedirect.com/science/article/pii/S0167642318301503}, \href {https://doi.org/10.1016/j.scico.2019.01.006} {\path{doi:10.1016/j.scico.2019.01.006}}.

\bibitem{Qcomp2020}
Carlos~E. Budde, Arnd Hartmanns, Michaela Klauck, Jan Kret{\'{\i}}nsk{\'{y}}, David Parker, Tim Quatmann, Andrea Turrini, and Zhen Zhang.
\newblock On correctness, precision, and performance in quantitative verification - {QComp} 2020 competition report.
\newblock In {\em ISoLA {(4)}}, volume 12479 of {\em Lecture Notes in Computer Science}, pages 216--241. Springer, 2020.
\newblock \href {https://doi.org/10.1007/978-3-030-83723-5\_15} {\path{doi:10.1007/978-3-030-83723-5\_15}}.

\bibitem{Chellaboina2009}
Vijaysekhar Chellaboina, Sanjay~P Bhat, Wassim~M Haddad, and Dennis~S Bernstein.
\newblock Modeling and analysis of mass-action kinetics.
\newblock {\em IEEE Control Systems Magazine}, 29(4):60--78, 2009.

\bibitem{Czerwinski2020}
W.~Czerwi{\'n}ski and G.~Zetzsche.
\newblock An {{Approach}} to {{Regular Separability}} in {{Vector Addition Systems}}.
\newblock In {\em {{ACM International Conference Proceeding Series}}}, pages 341--354. {Association for Computing Machinery}, July 2020.
\newblock \href {https://doi.org/10.1145/3373718.3394776} {\path{doi:10.1145/3373718.3394776}}.

\bibitem{Czerwinski2020b}
Wojciech Czerwi{\'n}ski, S{\l}awomir Lasota, Ranko Lazi{\'c}, {\relax J{\'e}}r{\^o}me Leroux, and Filip Mazowiecki.
\newblock The {{Reachability Problem}} for {{Petri Nets Is Not Elementary}}.
\newblock {\em Journal of the ACM}, 68(1):7:1--7:28, December 2020.
\newblock \href {https://doi.org/10.1145/3422822} {\path{doi:10.1145/3422822}}.

\bibitem{Czerwinski2021}
Wojciech Czerwi{\'n}ski and {\L}ukasz Orlikowski.
\newblock Reachability in vector addition systems is ackermann-complete.
\newblock In {\em 2021 IEEE 62nd Annual Symposium on Foundations of Computer Science (FOCS)}, pages 1229--1240, 2022.
\newblock \href {https://doi.org/10.1109/FOCS52979.2021.00120} {\path{doi:10.1109/FOCS52979.2021.00120}}.

\bibitem{Daigle2011}
Bernie J~Jr Daigle, Min~K Roh, Dan~T Gillespie, and Linda~R Petzold.
\newblock Automated estimation of rare event probabilities in biochemical systems.
\newblock {\em J Chem Phys}, 134(4):044110, Jan 2011.
\newblock \href {https://doi.org/10.1063/1.3522769} {\path{doi:10.1063/1.3522769}}.

\bibitem{Donovan2013}
Rory~M. Donovan, Andrew~J. Sedgewick, James~R. Faeder, and Daniel~M. Zuckerman.
\newblock Efficient stochastic simulation of chemical kinetics networks using a weighted ensemble of trajectories.
\newblock {\em The Journal of Chemical Physics}, 139(11):115105, September 2013.
\newblock \href {https://doi.org/10.1063/1.4821167} {\path{doi:10.1063/1.4821167}}.

\bibitem{Drawert2010}
Brian Drawert, Michael~J. Lawson, Linda Petzold, and Mustafa Khammash.
\newblock The diffusive finite state projection algorithm for efficient simulation of the stochastic reaction-diffusion master equation.
\newblock {\em The Journal of Chemical Physics}, 132(7):074101, 2010.
\newblock \href {https://arxiv.org/abs/https://doi.org/10.1063/1.3310809} {\path{arXiv:https://doi.org/10.1063/1.3310809}}, \href {https://doi.org/10.1063/1.3310809} {\path{doi:10.1063/1.3310809}}.

\bibitem{dresdeninverse}
Arnold Dresden.
\newblock {The fourteenth western meeting of the American Mathematical Society}.
\newblock {\em Bulletin of the American Mathematical Society}, 26(9):385 -- 396, 1920.

\bibitem{friedenberg2022probabilistic}
Meir~D Friedenberg, Adrian Lita, Mark~R Gilbert, Mioara Larion, and Orieta Celiku.
\newblock Probabilistic model checking of cancer metabolism.
\newblock {\em Scientific Reports}, 12(1):18870, 2022.

\bibitem{gardner2000construction}
Timothy~S Gardner, Charles~R Cantor, and James~J Collins.
\newblock Construction of a genetic toggle switch in escherichia coli.
\newblock {\em Nature}, 403(6767):339--342, 2000.

\bibitem{geva2006p53}
Naama Geva‐Zatorsky, Nitzan Rosenfeld, Shalev Itzkovitz, Ron Milo, Alex Sigal, Erez Dekel, Talia Yarnitzky, Yuvalal Liron, Paz Polak, Galit Lahav, and Uri Alon.
\newblock Oscillations and variability in the p53 system.
\newblock {\em Molecular Systems Biology}, 2(1):2006.0033, 2006.
\newblock URL: \url{https://www.embopress.org/doi/abs/10.1038/msb4100068}, \href {https://doi.org/10.1038/msb4100068} {\path{doi:10.1038/msb4100068}}.

\bibitem{Gillespie2009}
Dan~T. Gillespie, Min Roh, and Linda~R. Petzold.
\newblock Refining the weighted stochastic simulation algorithm.
\newblock In {\em Journal of Chemical Physics}, volume 130, 2009.

\bibitem{Hahn2009}
Ernst~Moritz Hahn, Holger Hermanns, Bj\"{o}rn Wachter, and Lijun Zhang.
\newblock Infamy: An infinite-state markov model checker.
\newblock In {\em Proceedings of the 21st International Conference on Computer Aided Verification}, CAV '09, pages 641--647, Berlin, Heidelberg, 2009. Springer-Verlag.

\bibitem{Han2007a}
Tingting Han and Joost-Pieter Katoen.
\newblock Counterexamples in probabilistic model checking.
\newblock In {\em International Conference on Tools and Algorithms for the Construction and Analysis of Systems}, pages 72--86. Springer, 2007.

\bibitem{Han2007}
Tingting Han and Joost-Pieter Katoen.
\newblock Providing {{Evidence}} of {{Likely Being}} on {{Time}}: {{Counterexample Generation}} for {{CTMC Model Checking}}.
\newblock In Kedar~S. Namjoshi, Tomohiro Yoneda, Teruo Higashino, and Yoshio Okamura, editors, {\em Automated {{Technology}} for {{Verification}} and {{Analysis}}}, volume 4762, pages 331--346. {Springer Berlin Heidelberg}, {Berlin, Heidelberg}, 2007.
\newblock \href {https://doi.org/10.1007/978-3-540-75596-8_24} {\path{doi:10.1007/978-3-540-75596-8_24}}.

\bibitem{harris2020array}
Charles~R. Harris, K.~Jarrod Millman, St{\'{e}}fan~J. van~der Walt, Ralf Gommers, Pauli Virtanen, David Cournapeau, Eric Wieser, Julian Taylor, Sebastian Berg, Nathaniel~J. Smith, Robert Kern, Matti Picus, Stephan Hoyer, Marten~H. van Kerkwijk, Matthew Brett, Allan Haldane, Jaime~Fern{\'{a}}ndez del R{\'{i}}o, Mark Wiebe, Pearu Peterson, Pierre G{\'{e}}rard-Marchant, Kevin Sheppard, Tyler Reddy, Warren Weckesser, Hameer Abbasi, Christoph Gohlke, and Travis~E. Oliphant.
\newblock Array programming with {NumPy}.
\newblock {\em Nature}, 585(7825):357--362, September 2020.
\newblock \href {https://doi.org/10.1038/s41586-020-2649-2} {\path{doi:10.1038/s41586-020-2649-2}}.

\bibitem{Hartmanns2014}
Arnd Hartmanns and Holger Hermanns.
\newblock The {M}odest {T}oolset: An integrated environment for quantitative modelling and verification.
\newblock In Erika {\'{A}}brah{\'{a}}m and Klaus Havelund, editors, {\em {TACAS}}, volume 8413 of {\em LNCS}, pages 593--598. Springer, 2014.

\bibitem{Hensel2021}
Christian Hensel, Sebastian Junges, Joost-Pieter Katoen, Tim Quatmann, and Matthias Volk.
\newblock The probabilistic model checker {{Storm}}.
\newblock {\em International Journal on Software Tools for Technology Transfer}, July 2021.
\newblock \href {https://doi.org/10.1007/s10009-021-00633-z} {\path{doi:10.1007/s10009-021-00633-z}}.

\bibitem{Hensel2022}
Christian Hensel, Sebastian Junges, Joost-Pieter Katoen, Tim Quatmann, and Matthias Volk.
\newblock The probabilistic model checker {{Storm}}.
\newblock {\em International Journal on Software Tools for Technology Transfer}, 24(4):589--610, August 2022.
\newblock \href {https://doi.org/10.1007/s10009-021-00633-z} {\path{doi:10.1007/s10009-021-00633-z}}.

\bibitem{hermanns1999multi}
Holger Hermanns, Joachim Meyer-Kayser, and Markus Siegle.
\newblock Multi terminal binary decision diagrams to represent and analyse continuous time markov chains.
\newblock In {\em Proc. NSMC}, volume~99, pages 188--207. Citeseer, 1999.

\bibitem{Israelsen2023}
Bryant Israelsen, Landon Taylor, and Zhen Zhang.
\newblock Efficient trace generation for rare-event analysis in chemical reaction networks.
\newblock In Georgiana Caltais and Christian Schilling, editors, {\em Model Checking Software}, pages 83--102, Cham, 2023. Springer Nature Switzerland.

\bibitem{Jegourel2012}
Cyrille Jegourel, Axel Legay, and Sean Sedwards.
\newblock Cross-entropy optimisation of importance sampling parameters for statistical model checking.
\newblock In {\em Proceedings of the 24th international conference on Computer Aided Verification}, CAV'12, pages 327--342, Berlin, Heidelberg, 2012. Springer-Verlag.
\newblock URL: \url{http://dx.doi.org/10.1007/978-3-642-31424-7_26}, \href {https://doi.org/10.1007/978-3-642-31424-7_26} {\path{doi:10.1007/978-3-642-31424-7_26}}.

\bibitem{Jegourel2013}
Cyrille Jegourel, Axel Legay, and Sean Sedwards.
\newblock Importance splitting for statistical model checking rare properties.
\newblock In Natasha Sharygina and Helmut Veith, editors, {\em Computer Aided Verification}, pages 576--591, Berlin, Heidelberg, 2013. Springer Berlin Heidelberg.

\bibitem{Jeppson2023}
Joshua Jeppson, Matthias Volk, Bryant Israelsen, Riley Roberts, Andrew Williams, Lukas Buecherl, Chris~J. Myers, Hao Zheng, Chris Winstead, and Zhen Zhang.
\newblock {STAMINA} in {C++:} modernizing an infinite-state probabilistic model checker.
\newblock In Nils Jansen and Mirco Tribastone, editors, {\em Quantitative Evaluation of Systems - 20th International Conference, {QEST} 2023, Antwerp, Belgium, September 20-22, 2023, Proceedings}, volume 14287 of {\em Lecture Notes in Computer Science}, pages 101--109. Springer, 2023.
\newblock \href {https://doi.org/10.1007/978-3-031-43835-6\_7} {\path{doi:10.1007/978-3-031-43835-6\_7}}.

\bibitem{Kahn1953}
H.~Kahn and A.~W. Marshall.
\newblock Methods of reducing sample size in monte carlo computations.
\newblock {\em Journal of the Operations Research Society of America}, 1(5):263--278, 1953.
\newblock \href {https://arxiv.org/abs/https://doi.org/10.1287/opre.1.5.263} {\path{arXiv:https://doi.org/10.1287/opre.1.5.263}}, \href {https://doi.org/10.1287/opre.1.5.263} {\path{doi:10.1287/opre.1.5.263}}.

\bibitem{Kahn1951}
Herman Kahn and Theodore~E Harris.
\newblock Estimation of particle transmission by random sampling.
\newblock {\em National Bureau of Standards applied mathematics series}, 12:27--30, 1951.

\bibitem{Karp1969}
Richard~M. Karp and Raymond~E. Miller.
\newblock Parallel program schemata.
\newblock {\em Journal of Computer and System Sciences}, 3(2):147--195, 1969.
\newblock URL: \url{https://www.sciencedirect.com/science/article/pii/S0022000069800115}, \href {https://doi.org/10.1016/S0022-0000(69)80011-5} {\path{doi:10.1016/S0022-0000(69)80011-5}}.

\bibitem{Kearns2005}
Daniel~B. Kearns and Richard Losick.
\newblock Cell population heterogeneity during growth of bacillus subtilis.
\newblock {\em Genes \& development}, 19 24:3083--94, 2005.

\bibitem{Kuwahara2008}
Hiroyuki Kuwahara and Ivan Mura.
\newblock An efficient and exact stochastic simulation method to analyze rare events in biochemical systems.
\newblock {\em The Journal of Chemical Physics}, 129(16):165101, October 2008.
\newblock \href {https://doi.org/10.1063/1.2987701} {\path{doi:10.1063/1.2987701}}.

\bibitem{Kwiatkowsa2012}
M.~Kwiatkowsa, G.~Norman, and D.~Parker.
\newblock The {PRISM} benchmark suite.
\newblock In {\em Quantitative Evaluation of Systems, International Conference on(QEST)}, volume~00, pages 203--204, 09 2012.
\newblock URL: \url{doi.ieeecomputersociety.org/10.1109/QEST.2012.14}, \href {https://doi.org/10.1109/QEST.2012.14} {\path{doi:10.1109/QEST.2012.14}}.

\bibitem{Kwiatkowska2007}
Marta Kwiatkowska, Gethin Norman, and David Parker.
\newblock {\em Stochastic Model Checking}, pages 220--270.
\newblock Springer Berlin Heidelberg, Berlin, Heidelberg, 2007.

\bibitem{Kwiatkowska2011}
Marta Kwiatkowska, Gethin Norman, and David Parker.
\newblock Prism 4.0: Verification of probabilistic real-time systems.
\newblock In {\em Proceedings of the 23rd International Conference on Computer Aided Verification}, CAV'11, pages 585--591, Berlin, Heidelberg, 2011. Springer-Verlag.

\bibitem{Legay2016}
Axel Legay, Sean Sedwards, and Louis-Marie Traonouez.
\newblock Rare {{Events}} for {{Statistical Model Checking}} an {{Overview}}.
\newblock In Kim~Guldstrand Larsen, Igor Potapov, and Ji{\v r}{\'\i} Srba, editors, {\em Reachability {{Problems}}}, Lecture {{Notes}} in {{Computer Science}}, pages 23--35. {Springer International Publishing}, 2016.
\newblock \href {https://doi.org/10.1007/978-3-319-45994-3_2} {\path{doi:10.1007/978-3-319-45994-3_2}}.

\bibitem{Leon2013}
Steven~J. Leon, {\AA}ke Bj{\"o}rck, and Walter Gander.
\newblock Gram-{{Schmidt}} orthogonalization: 100 years and more.
\newblock {\em Numerical Linear Algebra with Applications}, 20(3):492--532, 2013.
\newblock \href {https://doi.org/10.1002/nla.1839} {\path{doi:10.1002/nla.1839}}.

\bibitem{Leroux2011}
J\'{e}r\^{o}me Leroux.
\newblock Vector addition system reachability problem: a short self-contained proof.
\newblock In {\em Proceedings of the 38th Annual ACM SIGPLAN-SIGACT Symposium on Principles of Programming Languages}, POPL '11, pages 307--316, New York, NY, USA, 2011. Association for Computing Machinery.
\newblock \href {https://doi.org/10.1145/1926385.1926421} {\path{doi:10.1145/1926385.1926421}}.

\bibitem{leroux2012}
J{\'e}r{\^o}me Leroux.
\newblock {Vector Addition Systems Reachability Problem (A Simpler Solution)}.
\newblock In Andrei Voronkov, editor, {\em {Turing-100}}, volume~10 of {\em EPiC Series}, pages 214--228, Manchester, United Kingdom, June 2012. {Andrei Voronkov}.
\newblock URL: \url{https://hal.science/hal-00674970}.

\bibitem{Leroux2012VectorAS}
J{\'e}r{\^o}me Leroux.
\newblock Vector addition systems reachability problem (a simpler solution).
\newblock In {\em Turing-100}, 2012.
\newblock URL: \url{https://api.semanticscholar.org/CorpusID:2245739}.

\bibitem{Leroux2021}
J{\'e}r{\^o}me Leroux.
\newblock The reachability problem for petri nets is not primitive recursive.
\newblock In {\em 2021 IEEE 62nd Annual Symposium on Foundations of Computer Science (FOCS)}, pages 1241--1252, 2022.
\newblock \href {https://doi.org/10.1109/FOCS52979.2021.00121} {\path{doi:10.1109/FOCS52979.2021.00121}}.

\bibitem{Lipton1976}
R.J. Lipton.
\newblock {\em The reachability problem requires exponential space}.
\newblock Research report (Yale University. Department of Computer Science). Department of Computer Science, Yale University, 1976.
\newblock URL: \url{https://books.google.com/books?id=7iSbGwAACAAJ}.

\bibitem{madsen2014stochastic}
Curtis Madsen, Zhen Zhang, Nicholas Roehner, Chris Winstead, and Chris Myers.
\newblock Stochastic model checking of genetic circuits.
\newblock {\em ACM Journal on Emerging Technologies in Computing Systems (JETC)}, 11(3):1--21, 2014.

\bibitem{Mcmillan2019}
Kenneth~L. McMillan and Lenore~D. Zuck.
\newblock Compositional testing of internet protocols.
\newblock In {\em 2019 IEEE Cybersecurity Development (SecDev)}, pages 161--174, Sep. 2019.
\newblock \href {https://doi.org/10.1109/SecDev.2019.00031} {\path{doi:10.1109/SecDev.2019.00031}}.

\bibitem{Myers2009}
Chris~J. Myers.
\newblock {\em Engineering Genetic Circuits}.
\newblock Chapman \& Hall/CRC Mathematical and Computational Biology. Chapman \& Hall/CRC, 1 edition, July 2009.

\bibitem{Neupane2019}
Thakur Neupane, Zhen Zhang, Curtis Madsen, Hao Zheng, and Chris~J. Myers.
\newblock Approximation techniques for stochastic analysis of biological systems.
\newblock In Pietro Li{\`o} and Paolo Zuliani, editors, {\em Automated Reasoning for Systems Biology and Medicine}, volume~30, chapter~12, pages 327--348. Springer International Publishing, Cham, 2019.
\newblock \href {https://doi.org/10.1007/978-3-030-17297-8_12} {\path{doi:10.1007/978-3-030-17297-8_12}}.

\bibitem{Okamoto1959}
Masashi Okamoto.
\newblock Some inequalities relating to the partial sum of binomial probabilities.
\newblock {\em Annals of the Institute of Statistical Mathematics}, 10(1):29--35, 1959.
\newblock \href {https://doi.org/10.1007/BF02883985} {\path{doi:10.1007/BF02883985}}.

\bibitem{penrose1955}
R.~Penrose.
\newblock A generalized inverse for matrices.
\newblock {\em Mathematical Proceedings of the Cambridge Philosophical Society}, 51(3):406--413, 1955.
\newblock \href {https://doi.org/10.1017/S0305004100030401} {\path{doi:10.1017/S0305004100030401}}.

\bibitem{Rackoff1978}
Charles Rackoff.
\newblock The covering and boundedness problems for vector addition systems.
\newblock {\em Theoretical Computer Science}, 6(2):223--231, 1978.
\newblock URL: \url{https://linkinghub.elsevier.com/retrieve/pii/0304397578900361}, \href {https://doi.org/10.1016/0304-3975(78)90036-1} {\path{doi:10.1016/0304-3975(78)90036-1}}.

\bibitem{Roberts2022}
Riley Roberts, Thakur Neupane, Lukas Buecherl, Chris~J. Myers, and Zhen Zhang.
\newblock {STAMINA} 2.0: Improving scalability of infinite-state stochastic model checking.
\newblock In Bernd Finkbeiner and Thomas Wies, editors, {\em Verification, Model Checking, and Abstract Interpretation}, pages 319--331, Cham, 2022. Springer International Publishing.

\bibitem{Roh2011}
Min Roh, Bernie J.~Jr. Daigle, Dan~T. Gillespie, and Linda~R. Petzold.
\newblock State-dependent doubly weighted stochastic simulation algorithm for automatic characterization of stochastic biochemical rare events.
\newblock In {\em Journal of Chemical Physics}, volume 135. American Institute of Physics, 2011.

\bibitem{Roh2010}
Min Roh, Dan~T. Gillespie, and Linda~R. Petzold.
\newblock State-dependent biasing method for importance sampling in the weighted stochastic simulation algorithm.
\newblock In {\em Journal of Chemical Physics}, volume 133. American Institute of Physics, 2010.

\bibitem{Roh2016}
Min~K. Roh and Bernie~J. Daigle.
\newblock Sparse++: improved event-based stochastic parameter search.
\newblock {\em BMC Systems Biology}, 10(1):109, 2016.
\newblock \href {https://doi.org/10.1186/s12918-016-0367-z} {\path{doi:10.1186/s12918-016-0367-z}}.

\bibitem{Rosenbluth1955}
Marshall~N. Rosenbluth and Arianna~W. Rosenbluth.
\newblock Monte carlo calculation of the average extension of molecular chains.
\newblock {\em The Journal of Chemical Physics}, 23(2):356--359, 1955.
\newblock \href {https://arxiv.org/abs/https://doi.org/10.1063/1.1741967} {\path{arXiv:https://doi.org/10.1063/1.1741967}}, \href {https://doi.org/10.1063/1.1741967} {\path{doi:10.1063/1.1741967}}.

\bibitem{Soloveichik2010}
David Soloveichik, Georg Seelig, and Erik Winfree.
\newblock Dna as a universal substrate for chemical kinetics.
\newblock {\em Proceedings of the National Academy of Sciences}, 107(12):5393--5398, 2010.
\newblock URL: \url{https://www.pnas.org/doi/abs/10.1073/pnas.0909380107}, \href {https://arxiv.org/abs/https://www.pnas.org/doi/pdf/10.1073/pnas.0909380107} {\path{arXiv:https://www.pnas.org/doi/pdf/10.1073/pnas.0909380107}}, \href {https://doi.org/10.1073/pnas.0909380107} {\path{doi:10.1073/pnas.0909380107}}.

\bibitem{stewartBook}
William~J. Stewart.
\newblock {\em Iterative Methods}, pages 121--176.
\newblock Princeton University Press, 1994.
\newblock URL: \url{http://www.jstor.org/stable/j.ctv182jsw5.6}.

\bibitem{Taylor2023}
Landon Taylor, Bryant Israelsen, and Zhen Zhang.
\newblock Cycle and commute: Rare-event probability verification for chemical reaction networks.
\newblock In Alexander Nadel and Kristin~Y. Rozier, editors, {\em Proceedings of the 23rd Conference on Formal Methods in Computer-Aided Design -- FMCAD 2023}, pages 284--293. TU Wien Academic Press, 2023.
\newblock \href {https://doi.org/10.34727/2023/isbn.978-3-85448-060-0_37} {\path{doi:10.34727/2023/isbn.978-3-85448-060-0_37}}.

\bibitem{Villen-Altamirano2011}
Manuel Vill{\'e}n-Altamirano and Jos{\'e} Vill{\'e}n-Altamirano.
\newblock {\em The Rare Event Simulation Method RESTART: Efficiency Analysis and Guidelines for Its Application}, pages 509--547.
\newblock Springer Berlin Heidelberg, Berlin, Heidelberg, 2011.
\newblock URL: \url{http://dx.doi.org/10.1007/978-3-642-02742-0_22}, \href {https://doi.org/10.1007/978-3-642-02742-0_22} {\path{doi:10.1007/978-3-642-02742-0_22}}.

\bibitem{Wald1945}
A.~Wald.
\newblock Sequential tests of statistical hypotheses.
\newblock {\em The Annals of Mathematical Statistics}, 16(2):117--186, 1945.
\newblock URL: \url{http://www.jstor.org/stable/2235829}.

\bibitem{Wimmer2009}
R.~Wimmer, B.~Braitling, and B.~Becker.
\newblock {\em Counterexample Generation for Discrete-Time Markov Chains Using Bounded Model Checking}, volume 5403 LNCS of {\em Lecture {{Notes}} in {{Computer Science}}}.
\newblock January 2009.
\newblock \href {https://doi.org/10.1007/978-3-540-93900-9_29} {\path{doi:10.1007/978-3-540-93900-9_29}}.

\end{thebibliography}

\begin{appendix}
\newpage
    \section{Displacement Vector $f$}~\label{apx:offvec}

In this section, we present details in the calculation of $f$, introduced in Definition~\ref{def:indexedsub}. The requirements for $\vec{f}$ are that a) there exists some $\vec{x}$ such that $\vec{f} = M_0\vec{x} + \initSt$, which ensures that $\initSt \in S_0$ and b) $\initSt + \vec{f} \in \solutionSpace$, which ensures $S_i \cap \solutionSpace \neq \emptyset$. In order to ensure both of these conditions are met, we formulate the following two equations:

\begin{align*}
    \vec{f} &= M_0\vec{x} + \initSt \\
    \initSt + \vec{f} &= M_s \vec{y} + \st_p
\end{align*}

These equations can be rearranged as follows:

\begin{align*}
    \vec{f} &= M_s \vec{y} + \st_p - \initSt \\
    &= M_0\vec{x} + \initSt \\
    \vec{0} &= M_0\vec{x} + 2\initSt - M_s \vec{y} - \st_p \\
    \vec{0} &= \begin{bmatrix}
        M_0 & - M_s
    \end{bmatrix}
    \begin{bmatrix}
        \vec{x} \\ \vec{y}
    \end{bmatrix} + 2 \initSt - \st_p
\end{align*}

We have created a classical linear system which can be solved in a number of well-known ways for $\begin{bmatrix}
    \vec{x} \\ \vec{y}
\end{bmatrix}$. This result can then be used to find $\vec{f}$. In our implementation, we use linear least squares in \texttt{numpy}, which minimizes the magnitude of $\begin{bmatrix}
    \vec{x} \\ \vec{y}
\end{bmatrix}$. We found during our testing that the choice of $\vec{f}$, so long as it satisfies these properties, has little affect on the performance of \isr. In fact, even when a slightly \emph{incorrect} $\vec{f}$ is chosen (such as one where the linear system erroniously omits the $2\initSt$ term), \isr remains performant.
    \newpage
    \section{Description of Benchmarking Models}~\label{apx:models}

In this section, we describe the additional models used in our testing suite, taken from various sources including the \prism\ benchmark suite~\cite{Kwiatkowsa2012} and QComp~\cite{Qcomp2020}. The SSPD, EFC, MYP and SMR models constitute the original testing suite used by \ragtimer~\cite{Israelsen2023,Taylor2023}.

The \emph{Single Species Production-Degradation (SSPD) Model}
consists of two species in a two-reaction 
production-degradation interaction~\cite{Kuwahara2008}: 
\begin{align*}
    \react{1} &: \Sp_1 \xrightarrow{1.0} \Sp_1 + \Sp_2 \\
    \react{2} &: \Sp_2 \xrightarrow{0.025} \emptyset
\end{align*}
The initial state (i.e., initial molecule counts) for species 
vector \ensuremath{(\Sp_1, \Sp_2)} is \ensuremath{\initSt = [1, 40]^T}. 
The desired CSL property of this model is
\ensuremath{\probOp_{=?}(\Diamond^{[0, 100]} \, \Sp_2 = 80)}. 

The \emph{Enzymatic Futile Cycle (EFC)} as described by~\cite{Kuwahara2008} is shown below. The initial state for species vector 
\ensuremath{[\Sp_1, \Sp_2, \Sp_3, \Sp_4, \Sp_5, \Sp_6]} 
is \ensuremath{\initSt = [1, 50, 0, 1, 50, 0]^T}.
The rare-event property of interest is 
\ensuremath{\probOp_{=?}(\Diamond^{[0, 100]} \, \Sp_5 = 25)}.
\[
	\begin{array}{lll}
		\react{1} : \ \Sp_1 + \Sp_2 \xrightarrow{1.0} \Sp_3, &
		\react{2} : \ \Sp_3 \xrightarrow{1.0} \Sp_1 + \Sp_2, &
		\react{3} : \ \Sp_3 \xrightarrow{0.1} \Sp_1 + \Sp_5, ~~~ \\
		\react{4} : \ \Sp_4 + \Sp_5 \xrightarrow{1.0} \Sp_6, &
		\react{5} : \ \Sp_6 \xrightarrow{1.0} \Sp_4 + \Sp_5,~~~ &
		\react{6} : \ \Sp_6 \xrightarrow{0.1} \Sp_4 + \Sp_2.
	\end{array}
\]
Additionally, the \emph{Simplified Motility Regulation (SMR) Model} as described by~\cite{Kearns2005} is shown. The initial state for species vector 
[codY, flache, SigD\_hag, CodY, CodY\_flache, hag,CodY\_hag, 
SigD, Hag] is 
\ensuremath{\initSt = [1, 1, 1, 10, 1, 1, 1, 10, 10]^T}. 
The rare-event property is 	\ensuremath{\probOp_{=?}(\Diamond^{[0, 10]} \, 
\textrm{CodY} = 20)}.
\[
	\begin{array}{ll}
		\react{1} : \ \textrm{codY}  \xrightarrow{0.1} \textrm{codY} + 		\textrm{CodY}, 
  &
		\react{2} : \ \textrm{CodY} \xrightarrow{0.0002} \emptyset , 
  \\
		\react{3} : \ \textrm{flache} \xrightarrow{1.0} \textrm{flache} + 
		\textrm{SigD}, 
  &
		\react{4} : \ \textrm{SigD} \xrightarrow{0.0002} \emptyset,
  \\
		\react{5} : \ \textrm{SigD\_hag} \xrightarrow{1.0} \textrm{SigD} + 
		\textrm{hag} + \textrm{Hag},
  &
		\react{6} : \ \textrm{Hag} \xrightarrow{0.0002} \emptyset,
  \\
		\react{7} : \ \textrm{SigD} + \textrm{hag}\xrightarrow{0.01} 
		\textrm{SigD\_hag},
  &
		\react{8} : \ \textrm{SigD\_hag} \xrightarrow{0.1} \textrm{SigD} + 
		\textrm{hag},
  \\
		\react{9} : \ \textrm{CodY} + \textrm{flache} \xrightarrow{0.02} 
		\textrm{CodY\_flache},
  &
		\react{10} :  \textrm{CodY\_flache} \xrightarrow{0.1} \textrm{CodY} 
		+ \textrm{flache},
  \\
		\react{11} :  \textrm{CodY} + \textrm{hag} \xrightarrow{0.01} 
		\textrm{CodY\_hag},
  &
		\react{12} :  \textrm{CodY\_hag} \xrightarrow{0.1} \textrm{CodY} + 
		\textrm{hag}.\\
	\end{array}
\]

The \emph{Cancer Metabolism}\cite{friedenberg2022probabilistic} model simulates glycolysis and the tricarboxylic acid (TCA) cycle in cancer metabolism, incorporating both wild-type and mutant forms of isocitrate dehydrogenase (IDH), which catalyze key reactions involving tumor metabolites. Reaction rates in the model are instantiated using experimentally derived metabolic flux data measured under normoxic and hypoxic conditions, as well as a combination of these experimental rates with enzyme activity levels estimated from patient-derived mRNA expression data.

The \emph{Speed Independent Design} and \emph{Majority Gate Design}\cite{madsen2014stochastic}  are two types of genetic circuit architectures used to implement the Muller C-element. The Muller C-element is a critical state-holding logic gate commonly used in asynchronous circuits to coordinate the state of parallel processes. It exhibits a characteristic three-state logic: the output goes high when both inputs are high, goes low when both inputs are low, and retains its previous state when the inputs differ. In both genetic implementations, the inputs are IPTG and aTc, and the output is green fluorescent protein (GFP).

The \emph{Hill Toggle}\cite{gardner2000construction} is a gene regulatory network that captures the mutual repression dynamics between two proteins. In this network, protein 1 (p1) inhibits the expression of protein 2 (p2), and conversely, protein 2 (p2) inhibits protein 1 (p1). This reciprocal inhibition results in a bimodal steady-state distribution, enabling transient switching behavior between the two stable states.

This \emph{Tandem Queueing Network}~\cite{hermanns1999multi} model represents a two-stage tandem queueing network with a capacity of 511 customers per queue. Customers arrive at the first server node at a fixed rate, where they undergo a two-phase service process (each customer may experience one or two service phases) before proceeding to the second, single-phase server node. It also tracks the total number of customers present in the network at any given time.

The \emph{P53}~\cite{geva2006p53} model represents the oscillatory dynamics of the p53 regulatory network. It simulates the interactions among the tumor suppressor protein p53, Mdm2 mRNA (pMdm2), and the Mdm2 protein. 

\subsection{Properties}

All properties are of the form \cslEventually. For brevity, in the main body of the paper, we assigned short labels to each property rather than including their full CSL expressions. Table~\ref{tab:properties-csl} shows the CSL formulas for the properties labeled in Tables~\ref{tab:tool-benchmark-table} and~\ref{tab:verification-finally}.

\begin{table}[htbp]
  \centering
  \small
  \begin{tabular}{@{} cc @{}}
    \toprule
    \textbf{Property Lable} & \textbf{CSL Property} \\
    \midrule
    p\_CM1   & $\probOp_{=?}[\Diamond^{[0,15]}(\mathrm{Gluc}\geqslant 5)]$\\
    p\_CM2   & $\probOp_{=?}[\Diamond^{[0,15]}(\mathrm{Pyr}\geqslant 5)]$\\
    p\_CM3   & $\probOp_{=?}[\Diamond^{[0,15]}(\mathrm{Lip}\geqslant 5)]$\\
    p\_SSPD  & $\probOp_{=?}[\Diamond^{[0,100]}(S_{1}\geqslant 80)]$\\
    p\_EFC   & $\probOp_{=?}[\Diamond^{[0,100]}(S_{4}\leqslant 25)]$\\
    p\_MYP   & $\probOp_{=?}[\Diamond^{[0,20]}(S_{5}\geqslant 50)]$\\
    p\_SMR   & $\probOp_{=?}[\Diamond^{[0,10]}(S_{5}\geqslant 20)]$\\
    p\_P53\_1 & $\probOp_{=?}[\Diamond^{[0,5]}(P53\geqslant 120)]$\\
    p\_P53\_2 & $\probOp_{=?}[\Diamond^{[0,100]}(\mathrm{Mdm2}=0)]$\\
    p\_MGD   & $\probOp_{=?}[\Diamond^{[0,2100]}((EE\geqslant 40)\,\&\,(CC<20))]$\\
    p\_SID   & $\probOp_{=?}[\Diamond^{[0,2100]}((S_{2}\geqslant 80)\,\&\,(S_{3}<20))]$\\
    p\_HT1   & $\probOp_{=?}[\Diamond^{[0,10]}(p2>10)]$\\
    p\_HT2   & $\probOp_{=?}[\Diamond^{[0,10]}((p2>5)\,\&\,(p1>5))]$\\
    p\_TQN   & $\probOp_{=?}[\Diamond^{[0,0.2]}(sc=511)]$\\
    \bottomrule
  \end{tabular}
  \caption{CSL formulas for the properties tested in this paper.}
  \label{tab:properties-csl}
\end{table}
\end{appendix}

\appendix

\end{document}